\documentclass[prd,nofootinbib,twocolumn,superscriptaddress,preprintnumbers,balancelastpage,floatfix]{revtex4-1}
\usepackage[margin=1in]{geometry}

\usepackage{graphicx}
\graphicspath{{figures/}}
\usepackage{color}
\usepackage{url}
\usepackage{hyperref}
\hypersetup{
    colorlinks = true,
    citecolor  = red,
    linkcolor  = blue
}
\usepackage{physics} 
\usepackage{soul} 
\usepackage{array} 
\newcolumntype{C}[1]{>{\centering\arraybackslash}p{#1}}
\usepackage{fontawesome} 

\ifdefined\darkmode
\pagecolor[rgb]{0.7,0.7,0.7}

\fi

\usepackage{amsmath, amssymb}
\usepackage{halloweenmath}

\renewcommand{\dd}{\text{d}}

\newcommand{\figref}[1]{Fig.~\ref{#1}}
\renewcommand{\eqref}[1]{Eq.~(\ref{#1})}
\newcommand{\bpm}{\begin{pmatrix}}
\newcommand{\epm}{\end{pmatrix}}
\newcommand{\bbm}{\begin{bmatrix}}
\newcommand{\ebm}{\end{bmatrix}}

\definecolor{darkmagenta}{rgb}{0.55, 0.0, 0.55}
\definecolor{teal}{rgb}{0.0, 0.6, 0.6}
\definecolor{overleaf}{rgb}{0.0, 0.7, 0.0}

\newcommand{\gagg}{g_{a\gamma\gamma}}
\newcommand{\vx}{{\vec{x}}}
\newcommand{\hx}{{\hat{x}}}

\newcommand{\ourtitle}{Looking in the axion mirror: An all-sky analysis of stimulated decay}

\begin{document}
\preprint{MIT-CTP/5603}
\title{\ourtitle \vspace*{-0.3cm}}
\author{Yitian Sun}
\affiliation{MIT Center for Theoretical Physics, Massachusetts Institute of Technology, Cambridge, MA 02139, USA}
\author{Katelin Schutz}
\affiliation{Department of Physics, McGill University, Montr\'eal, QC H3A 2T8, Canada}
\affiliation{Trottier Space Institute at McGill, Montr\'eal, QC H3A 2T8, Canada}
\author{Harper Sewalls}
\affiliation{Department of Physics, McGill University, Montr\'eal, QC H3A 2T8, Canada}
\affiliation{Trottier Space Institute at McGill, Montr\'eal, QC H3A 2T8, Canada}
\affiliation{UC Berkeley Department of Physics, Berkeley, CA 94720, USA}
\author{Calvin Leung}
\affiliation{UC Berkeley Department of Astronomy, Berkeley, CA 94720, USA}
\affiliation{MIT Kavli Institute for Astrophysics and Space Research, Massachusetts Institute of Technology, Cambridge, MA 02139, USA}
\affiliation{Department of Physics, Massachusetts Institute of Technology, Cambridge, MA 02139, USA}
\affiliation{NASA Einstein Fellow}
\author{Kiyoshi Wesley Masui}
\affiliation{MIT Kavli Institute for Astrophysics and Space Research, Massachusetts Institute of Technology, Cambridge, MA 02139, USA}
\affiliation{Department of Physics, Massachusetts Institute of Technology, Cambridge, MA 02139, USA}

\begin{abstract} \noindent 
Axion dark matter (DM) produces echo images of bright radio sources via stimulated decay. These images appear as a faint radio line centered at half the axion mass, with the line width set by the DM velocity dispersion. Due to the kinematics of the decay, the echo can be emitted in the direction nearly opposite to the incoming source of stimulating radiation, meaning that axions effectively behave as imperfect monochromatic mirrors. We present an all-sky analysis of axion DM-induced echo images using extragalactic radio point sources, Galactic supernova remnants (SNRs), and Galactic synchrotron radiation (GSR) as sources of stimulating radiation. The aggregate signal strength is not significantly affected by unknown properties of individual sources of stimulating radiation, which we sample from an empirical distribution to generate an ensemble of realizations for the all-sky signal template. We perform forecasts for CHIME, HERA, CHORD, HIRAX, and BURSTT, finding that they can run as competitive axion experiments simultaneously with other objectives, requiring no new hardware.~\href{https://github.com/yitiansun/axion-mirror}{\faGithub}
\end{abstract}
\maketitle

\section{Introduction}
The nature of dark matter (DM) is one of the most significant outstanding questions in modern physics. Axions and axion-like particles are among the most widely pursued DM candidates due to the ubiquity of such particles in UV-complete theories~\cite{witten1984some, svrcek2006axions, Arvanitaki:2009fg, acharya2010m, Cicoli:2012sz} and their role in addressing several outstanding issues in the Standard Model, including the strong-$CP$ problem~\cite{peccei1977cp, weinberg1978new, wilczek1978problem, abbott1983cosmological, preskill1983cosmology, dine1983not} and the matter-antimatter asymmetry~\cite{Co:2020xlh}. Several dedicated axion DM detection experiments have been built or proposed~\cite{panfilis1987limits, wuensch1989results, hagmann1990results, Asztalos:2003px, Boutan:2018uoc, ADMX:2018gho, ADMX:2021nhd, ADMX:2019uok, HAYSTAC:2018rwy, HAYSTAC:2020kwv, Kahn:2016aff, Salemi:2021gck, Lawson:2019brd, Wooten:2022vpj, Caldwell:2016dcw, MADMAX:2019pub, BREAD:2021tpx, DMRadio:2022pkf, Berlin:2020vrk}, most of which rely on the conversion of axions to photons in an ambient electromagnetic field, which arises due to the $g_{a\gamma \gamma} a( \vec{E}\cdot \vec{B})$ interaction term appearing in the Lagrangian. A less well-studied consequence of this interaction is the decay of axions; notably, while the spontaneous decay rate is slow in unconstrained parts of axion parameter space, the stimulated decay rate can be significantly enhanced~\cite{arza2019production}. 

\begin{figure}[t]
    \centering
\includegraphics[width=0.48\textwidth]{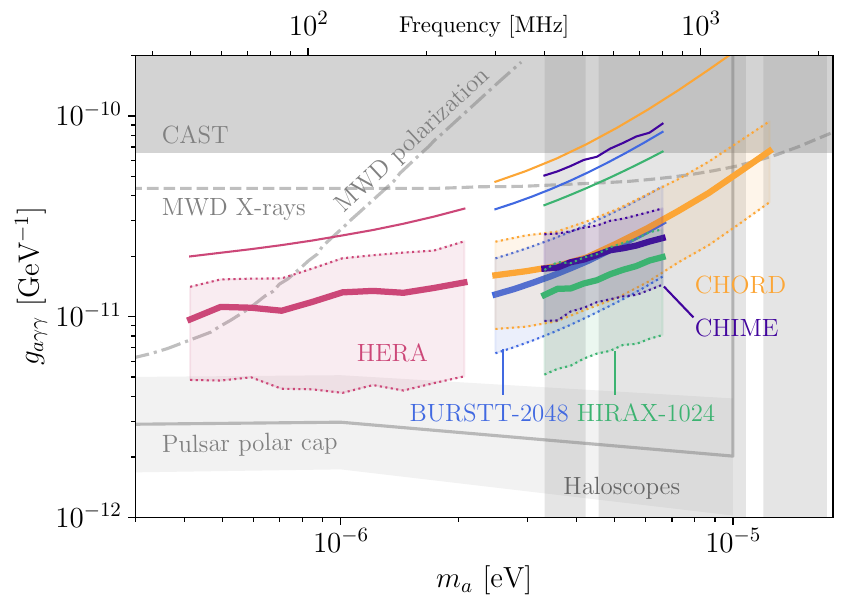}\vspace{-0.5cm}
    \caption{\textbf{Projected reach of various arrays operating as survey interferometers.} These instruments are sensitive to the cumulative signal of all bright radio sources that induce stimulated axion decay. We assume five years of integration time, comparable to the CHIME archival data. The bands show the 95\% containment of 300 realizations of the total signal, with SNRs producing the dominant contribution. We sample any unmeasured information about observed SNRs from empirical distributions, which gives rise to the statistical variation. The thin lines above the bands show sensitivity from only including stimulating radiation from GSR. Grey regions and lines correspond to existing limits~\cite{panfilis1987limits, wuensch1989results, hagmann1990results, Asztalos:2003px, Boutan:2018uoc, ADMX:2018gho, ADMX:2021nhd, ADMX:2019uok, HAYSTAC:2018rwy, HAYSTAC:2020kwv, Kahn:2016aff, Salemi:2021gck, Anastassopoulos:2017ftl, Dessert:2021bkv, Dessert:2022yqq, Noordhuis:2022ljw, ciaran_o_hare_2020_3932430}.}
    \vspace{-0.5cm}
    \label{fig:reach}
\end{figure}

Due to the non-relativistic kinematics of axion DM in the Galactic halo, DM axions undergoing stimulated decay tend to produce photons that are nearly back to back, with the decay axis pointing along the direction of the incoming radiation. Each photon has an energy corresponding to half the axion mass, with a $\sim10^{-3}$-level Doppler broadening due to the velocity dispersion in the DM halo. Recently, several studies have considered the feasibility of observing the ``echo'' image of axion decay induced by individual bright astrophysical sources, most notably Cygnus A~\cite{ghosh2020axion} and supernova remnants (SNRs)~\cite{sun2022axion, buen2022axion}. These individual sources would produce images antipodal to the source, obtained by integrating over all the axion decay in a DM column oriented along the line of sight. Ref.~\cite{sun2022axion} showed that SNR-stimulated axion decay could be observed with the world's most powerful existing radio telescopes like the Five-hundred-meter Aperture Spherical radio Telescope (FAST)~\cite{nan2011five}, potentially exploring axions in the $\sim 0.6~\mu$eV-30~$\mu$eV mass range at couplings below the limit set by the CERN Axion Solar Telescope (CAST)~\cite{Anastassopoulos:2017ftl}.

In this work, we instead consider radio telescopes such as the Canadian Hydrogen Intensity Mapping Experiment (CHIME)~\cite{amiri2022overview} and other existing and planned telescopes that survey a large fraction of the sky over a period of years. Our key result is summarized in Fig.~\ref{fig:reach}. We previously showed that survey telescopes are less sensitive than FAST for detecting stimulated axion decay from individual SNRs~\cite{sun2022axion}. However, we can improve the sensitivity of searches with survey telescopes by around two orders of magnitude in signal strength (or around an order of magnitude in axion coupling reach) by making use of the cumulative signal over the entire sky from all possible sources of stimulating radiation rather than only focusing on the strongest individual sources. 

The rest of this paper is organized as follows. In Section~\ref{sec:formalism}, we review the formalism for computing the stimulated decay flux. We include the flux of both the decay photons going in the direction of the incoming radiation (``forwardschein'') and the photons going in nearly the opposite direction (``gegenschein''), including for the first time the contribution from sources that are directly in front of their gegenschein images. In Section~\ref{sec:source}, we describe how we model the most important sources of stimulating radiation including radio galaxies and quasars, SNRs, and Galactic synchrotron radiation (GSR). SNRs have the largest uncertainties on their individual contributions to the axion decay signal due to theoretical uncertainties in their luminosity during the free-expansion phase of evolution. We discuss how we use empirical distributions to generate several realizations for the properties of both observed SNRs with incomplete information and also the ``SNR graveyard'' of SNRs that are too faint to currently be detected~\cite{buen2022axion}. In Section~\ref{sec:sensitivity}, we then describe our prescription for forecasting the sensitivity of various radio telescopes to axion decay. Outlook and concluding remarks follow in Section~\ref{sec:conclusions}.

\section{Stimulated decay intensities}
\label{sec:formalism}
\begin{figure}[t]
    \centering
    \includegraphics[clip, width=0.48\textwidth]{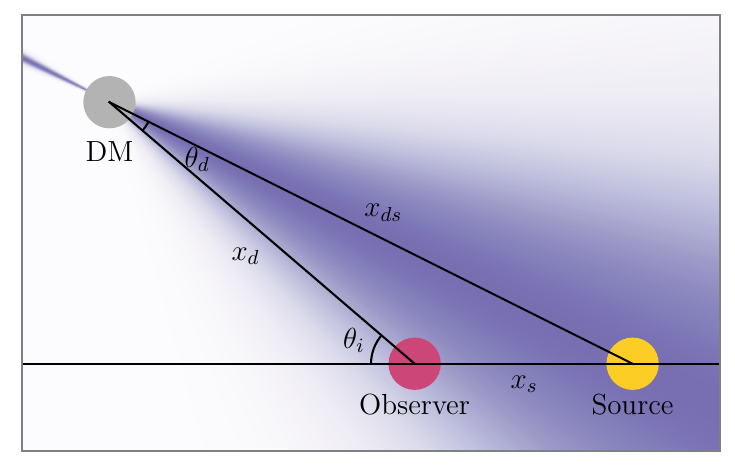}
    \caption{\textbf{Gegenschein from sources behind the observer.} Axion DM undergoes stimulated decay in the presence of source radio photons. The resulting decay photon distribution is shown in purple. The backward traveling (``gegenschein'') photon is smeared due to DM's velocity dispersion in the galactic halo. Consequently, gegenschein can be observed in directions that deviate slightly from the antipodal direction of the source. } \label{fig:gegenschein}
\end{figure}

The general setup is depicted in Figs.~\ref{fig:gegenschein}--\ref{fig:forwardschein}, where the observer is at the origin. We use $\vx_s$ to denote the vector from the origin to the source volume element, $\vx_d$ to describe the vector to the DM volume element, and the distance between the DM and source elements is $\vx_{ds} \equiv \vx_d-\vx_s$. We consider general source geometries, from point sources to sources that are spatially extended such as GSR, described in Section \ref{sec:gsr}. We take the source to have a specific volume emissivity $j_\nu(\vx_s)$, which is the luminosity per unit source volume at frequency $\nu$ at location $\vx_s$. The total specific luminosity of the source is
\begin{equation}
    L_\nu(t) = \int_\text{source} j_\nu(\vx_s, t) d^3 x_s
\end{equation}
and one can recover the luminosity of a point source with an emissivity that is a delta function at the source location.

As seen by DM at location $\vx_d$, the specific intensity coming from a source emitting isotropically will be
\begin{equation} \label{eq:I_nu_source}
    I_\nu(\hx_{ds}) = \int \frac{j_\nu(\vx_d-\vx_{ds})}{4\pi} \dd x_{ds},
\end{equation}
where $\hx_{ds}$ denotes the unit vector pointing in the direction of $\vx_{ds}$, and $\dd x_{ds}$ is the one dimensional line-of-sight integral along a fixed direction $\hx_{ds}$ (with fixed $\vx_d$ and varying $\vx_s$).

Due to the incoming flux of radiation that stimulates axions to decay, there will be observable emission coming from the DM to the observer. There are two outgoing photons per decay, with one photon in the same direction as the incoming radiation and the second photon emitted in the direction that conserves energy and momentum. If the axions are perfectly at rest with respect to the source, then the two photons will be back to back. However, in general axion DM has some velocity dispersion, so the forward emission is still in the direction of $\vx_{ds}$ whereas the backwards emission has some angular distribution due to the transverse boost from the axion frame to the source frame. Depending on the relative configuration of the observer, DM and source, the emission that is ultimately observed will either be the collimated forward emission or the smeared backward emission. To account for these two emission modes, we write the distribution of the emitted photons in terms of a decay kernel
\begin{equation} \label{eq:Fie}
    \mathcal{F}(\hat i,\hat e) = \delta^2(\hat i - \hat e) + f(\hat i,\hat e),
\end{equation}
where $\hat i$ stands for the incoming photon direction (equal to $\hx_{ds}$ in our setup) and $\hat e$ represents the emitted photon direction. Here $\delta^2$ is a 2D delta function on the unit sphere corresponding to the decay photon going in the original direction that we will refer to as ``forwardschein'', and $f$ is the smearing envelope of the ``gegenschein'' photon determined by the transverse DM velocity distribution. We normalize $f$ such that its integral over the unit sphere is 1, and $\mathcal{F}$ would normalize to 2, corresponding to the emitted photon number. In our analysis, we take the transverse DM velocities to be Gaussian-distributed with a characteristic dispersion $\sigma_d$. For DM with transverse velocity $v\ll1$ (we work in units where $c = \hbar = 1$ throughout this entire paper), the echo photon must make an angle of $2v$ with respect to the incoming photon direction in order to conserve momentum. This means that $f$ takes the form a Gaussian with angular dispersion $\theta_d = 2 \sigma_d$.

We take $\sigma_d =$ 116~km/s$~\sim0.4\times10^{-3}$ in the local Milky Way halo, which gives a similar velocity distribution as the one inferred from indirect measurement~\cite{necib2018inferred}. We additionally take the DM density $\rho$ to be distributed as a Navarro-Frenk-White profile with scale radius $r_s=16$~kpc \cite{nitschai2020first} and local DM density at $r_\odot=8.2$~kpc \cite{du2018detection, 2019A&A...625L..10G, abuter2020arxiv} of $0.44$~GeV/cm$^3$ \cite{nitschai2020first}. Note that in the context of such a density profile, the velocity dispersion will peak at intermediate radii, near the scale radius~\cite{Hoeft:2003ea}. Therefore, the velocity dispersion adopted in this work is likely an overestimate of the true velocity dispersion in the inner Milky Way DM halo, where the density and stimulated decay are enhanced. Given the difficulties in determining even the local DM velocity distribution, we assume that the local velocity dispersion is representative of the entire Milky Way. This assumption is conservative because having a larger dispersion only smears the signal, as described further below. We do not take into account any aberration effects, \emph{i.e.} blurring of the source due to the bulk relative motion of the source, Earth, and DM halo. Aberration effects are in general smaller than the blurring due to velocity dispersion, as was shown explicitly for the case of SNRs in Ref.~\cite{buen2022axion}.

We can use the photon decay kernel to determine total intensity for stimulated decay,
\begin{equation} \label{eq:Ist}
    \begin{aligned}
        I_\text{st}(-\hx_d,t)=&\frac{\gagg^2}{16}\iint\frac{\dd x_d\dd^3\vx_s}{4\pi x_{ds}^2}\rho(\vx_d) \\
        &j_\nu(\vx_s,t-x_d-x_{ds})\mathcal{F}(\hx_{ds}, -\hx_d),
    \end{aligned}
\end{equation}
where we have identified the emitted photon direction as $\hat e=-\hx_d$ and where the 1D DM integral is along the $\hx_d$ direction. See Appendix \ref{appd:Ist} for a detailed derivation. We can see that the stimulated intensity is proportional to the (time delayed) source emissivity, and involves a column integral of the DM density, but is complicated by the dependence on $x_{ds}$ and $\mathcal F$. To gain a clearer physical interpretation of this expression, we separately consider three relative configurations of the source, the DM column, and the observer in the following Subsections, and discuss the limiting case of point sources in each section.

\subsection{Gegenschein for sources behind the observer}
We first consider the already well-studied case of gegenschein where the source for the stimulating photon is ``behind'' an observer looking in the direction of the decaying DM column, as depicted in \figref{fig:gegenschein}. In this configuration, only the backwards portion of the decay kernel, $f(\hat i, \hat e)$, is relevant. For a source volume element (or a point source), this means the gegenschein image deviates from the antipodal direction of the source element by an angle $\theta_{i}$ such that
\begin{equation} \label{eq:sin_theta_i}
    \sin\theta_i=\sin\theta_d\cdot x_{ds}/x_s,
\end{equation}
where $\theta_d$ is the angle between the echo photon and the stimulating photon. For a source volume element $\dd^3\vx_s=x_s^2\dd x_s\dd\Omega_s$, we can then rewrite \eqref{eq:Ist} as
\begin{equation} \label{eq:Ig}
     \begin{aligned}
        I_\text{g}(-\hx_d,t)=&\frac{\gagg^2}{16}\iint\dd x_d\dd\Omega_s\dd x_s~\rho(\vx_d) \\
        &\times\frac{1}{4\pi}j_\nu(\vx_s, t-x_d-x_{ds})h(\theta_i)
    \end{aligned}
\end{equation}
where we have defined
\begin{equation} \label{eq:h_theta_i}
    h(\theta_i)\equiv \frac{x_s^2}{x_{ds}^2}f\left(\arcsin\left(\frac{x_s}{x_{ds}}\sin\theta_i\right)\right),
\end{equation}
where the argument of $f$ comes from inverting \eqref{eq:sin_theta_i}. 

When $x_{ds}/x_s\ll\theta_d^{-1} \approx 10^3$, $\theta_i$ is small and $\theta_i=\theta_d\cdot x_{ds}/x_s$. In that limit, we can see that $h(\theta_i)$ is just a scaled version of the $f(\theta_d)$ distribution, $h(\theta_i) \approx\frac{x_s^2}{x_{ds}^2}f\left(\frac{x_s}{x_{ds}}\theta_i\right)$. For a Gaussian distribution $f$, $h$ has the same normalization in the small-angle limit, $\int h(\theta_i)\dd\Omega_i=\int f(\theta_d)\dd\Omega_d=1$. We thus call $x_{ds}/x_s\lesssim \theta_d^{-1}$ the \emph{focused limit} in which the overall power of the gegenschein image is still the same as in the case where there is no DM velocity dispersion. In the opposite limit when $x_{ds}/x_s$ is large, most gegenschein photons are deflected away from the observer, causing the overall power of gegenschein to diminish.

\eqref{eq:Ig} has a simple physical interpretation: for a given source volume element defined by $\dd x_s\dd\Omega_s$, the gegenschein intensity it induces is the integral of the DM density along the DM column, weighted by the $h(\theta_i)$ distribution (which depends on $x_d$). For a thin isotropic source with negligible depth, one can take
\begin{equation}
    j_\nu(\vx_s,t)=4\pi I_\nu(t+x_s) \delta(x_s-x_{s0}),
\end{equation}
where $I_\nu(t)$ is the observed specific intensity on Earth and where the $x_s$ term in the argument accounts for the retarded time. If one considers point sources with
\begin{equation}
    I_\nu(t)=S_\nu(t)\delta^2(\hx_s-\hx_{s0})
\end{equation}
for specific flux $S_\nu$, one can arrive at the gegenschein intensity for point sources in the focused limit,
\begin{equation} \label{eq:Ig_ps}
    I_\text{g}(-\hx_d,t)=\frac{\gagg^2}{16}\int\dd x_d~\rho(x_d)S_\nu(t-2x_d)h(\theta_i),
\end{equation}
where we have used $x_d+x_{ds}-x_s\approx2x_d$ in the focused limit. From this expression one can readily recover Eq.~2 in \cite{sun2022axion}
by integrating over the solid angle $\Omega_i$ corresponding to $\theta_i$.

\subsection{Gegenschein from sources in front of the observer}
\begin{figure}[t]
    \centering
    \includegraphics[clip, width=0.48\textwidth]{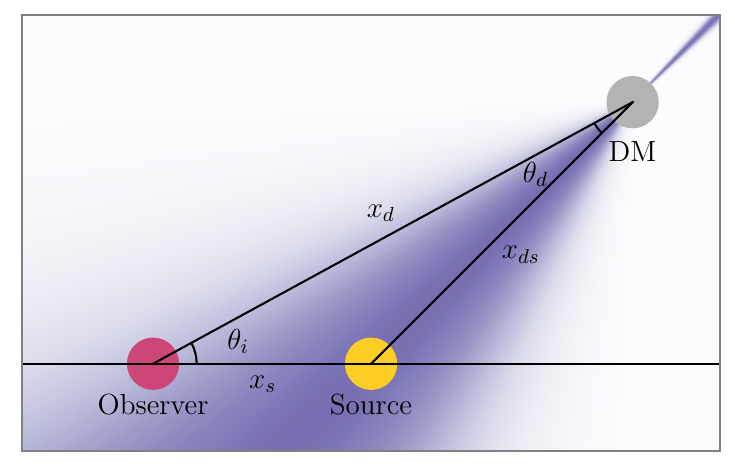}
    \caption{\textbf{Gegenschein from sources in front of the observer.} The axion DM in front of the observer and behind the source undergoes stimulated decay, producing a smeared gegenschein radiation and collimated forward-going radiation.}
    \label{fig:front_gegenschein}
\end{figure}
In the previous Subsection, we derived a straightforward extension of gegenschein considered in Refs.~\cite{ghosh2020axion, sun2022axion, buen2022axion}. In this Subsection and the following one, we will consider different source-observer configurations where the image and the source directions are roughly aligned. Na\"ively, this geometry may seem less optimal than the classic gegenschein geometry, as the decay photons come from roughly the same direction as source photons and thus the source itself may pose a significant background that overwhelms the faint radio line coming from axion decay. Nevertheless, our analysis reveals that for the sources we consider, described further in Section~\ref{sec:source}, these same-direction components of stimulated decay are important. We provide a brief rationale here and expand on this point further when we discuss the sources in depth in Section~\ref{sec:source}.

Heuristically, diffuse GSR is the main background for the classical gegenschein geometry from any source. Therefore, there is no reason not to consider geometries where the image and source are aligned when the source itself is GSR. In other words, the backgrounds coming from the source direction are similar to the backgrounds antipodal to the source direction when the source is GSR. Meanwhile, for time-varying point sources of stimulating radiation (most notably SNRs), the primary contribution to the stimulated decay brightness comes from the earliest stages of SNR emission. As observed presently, SNRs are orders of magnitude dimmer than they would have been when stimulating the decay. Consequently, the current brightness of the remnants does not necessarily overwhelm the stimulated decay photons arriving from approximately the same direction. This is especially relevant given the possibility that the stimulated decay image could be smeared by transverse DM velocities and would appear more spatially extended compared to the point source. 

Having established some heuristic arguments for why stimulated decay photons from the source direction can contribute appreciably to the signal, we first consider the case where the DM element is in the source direction behind the source from the observer perspective. This configuration is depicted in \figref{fig:front_gegenschein}. In this configuration, we would still observe the decay photon coming from the direction opposite to that of the stimulating photon, so this is still an example of gegenschein. We distinguish this case from the case considered in the previous Subsection by referring to it as ``front gegenschein". We note that \eqref{eq:sin_theta_i} still holds for front gegenschein, and the focused (small image-angle) limit still applies for most relevant DM-source distances $x_{ds}\ll \theta_d^{-1} x_s$. The front gegenschein intensity is therefore identical to that of the classic gegenschein configuration,
\begin{equation} \label{eq:Ifg}
     \begin{aligned}
        I_\text{fg}(-\hx_d,t)=&\frac{\gagg^2}{16}\iint\dd x_d\dd\Omega_s\dd x_s~\rho(\vx_d) \\
        &\times\frac{1}{4\pi}j_\nu(\vx_s, t-x_d-x_{ds})h(\theta_i)
    \end{aligned}
\end{equation}
where $h$ is defined the same way as in \eqref{eq:h_theta_i}. Note that unlike the classical gegenschein geometry, $x_{ds}$ is not always larger than $x_s$. For point sources, upon making the small image angle limit one can obtain a similar expression as \eqref{eq:Ig_ps},
\begin{equation} \label{eq:Ifg_ps}
    I_\text{fg}(-\hx_d,t)=\frac{\gagg^2}{16}\int\dd x_d~\rho(x_d)S_\nu(t-2x_{ds})h(\theta_i),
\end{equation}
where we have approximated $x_d+x_{ds}-x_s=2x_{ds}$ for approximately co-linear source and DM elements.

In the case of SNRs as sources,  most of the gegenschein signal originates from the stimulating radiation emitted during the earliest stages of SNR evolution. Therefore, for SNRs with age $t_0$, the most significant part of the DM column is a distance $\sim t_0/2$ away from the observer (taking the source to be relatively nearby). In other words, the peak signal comes from $x_d\sim t_0/2$ in the case of regular gegenschein, and $x_{ds}\sim t_0/2$ in the case of front gegenschein. For a time-varying point source like a SNR, \eqref{eq:sin_theta_i} therefore implies that the smearing effect is generally larger for regular gegenschein than it is for front gegenschein. On the other hand, the degree of smearing for an extended source depends on a double integral over the DM column and the source column. Therefore it is not possible to make a general comparison of the smearing effect between regular gegenschein and front gegenschein for diffuse sources.

\subsection{Forwardschein for point sources and extended sources}
\begin{figure}[t]
    \centering
    \includegraphics[clip, width=0.48\textwidth]{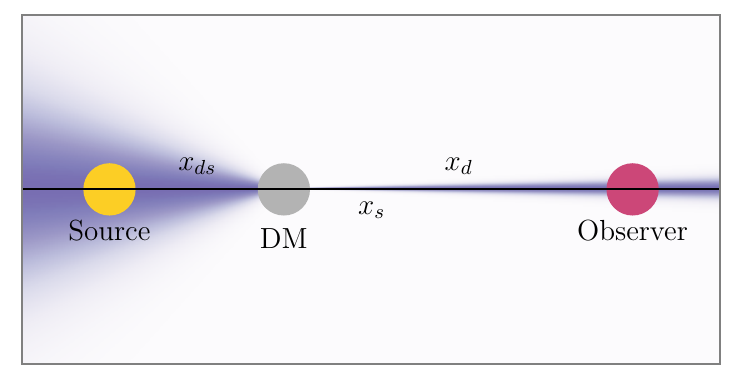}
    \caption{\textbf{Forwardschein from DM in between the source and observer.} The stimulated radiation moving towards the observer is collimated, so the DM must be exactly in between the source and the observer. Stimulating radiation from diffuse sources contributes much more significantly to the signal than radiation from SNRs, due to the lack of time delay between the arrival of stimulating radiation and the decay products. With this configuration, we are therefore unable to make use of the remnants' much brighter past emission.}
    \label{fig:forwardschein}
\end{figure}

Finally, we consider the case where the DM volume element is between the source and the observer, which we refer to as ``forwardschein''. While similar to front gegenschein where the image is in the same direction as the source, the key difference is that the forward decay photon must be in the same momentum state as the stimulating photon. Therefore, regardless of the velocity of the decaying DM, the forwardschein image lies directly on top of the source image. This is equivalent to taking only the delta function term in \eqref{eq:Fie} as being observable. In this configuration, the stimulated decay intensity \eqref{eq:Ist} can be simplified as
\begin{equation} \label{eq:If}
    I_\text{f}(-\hx_d,t)=\frac{\gagg^2}{16}\iint\dd x_d\dd x_s\rho(\vx_d)\frac{j_\nu(\vx_s,t-x_s)}{4\pi},
\end{equation}
where we have used the fact that
\begin{equation}
    \int\dd\Omega_s\frac{x_s^2}{x_{ds}^2}\delta^2(\hx_{ds}+\hx_d)=1,
\end{equation}
with an analogous normalization as the gegenschein distribution $h$ in \eqref{eq:h_theta_i}. Note that the discussion of the small angle limit for gegenschein's does not apply here, as the DM is always on the source LOS in order for its forwardschein to be observable. Note also that although the momentum of the decay photon is the same as the stimulating photon, the decay line is still broadened by the DM velocity dispersion due to the $m_a/2$ resonance occurring in the DM frame rather than in the observer's frame. For point sources, the above intensity can be further simplified as
\begin{equation}
    I_\text{f}(-\hx_d,t)=\frac{\gagg^2}{16}I_\nu(t)\int\dd x_d\rho(\vx_d),
\end{equation}
where the source intensity $I_\nu(t)$ can be factored out of the DM column integral. However, the contribution to the signal-to-noise from point source forwardschein is not expected to be strong: unlike for the case of gegenschein, there is no extra time delay between the arrival of the source photon and the decay photon, and we therefore cannot access past brightness history of point sources like SNRs. We will therefore mainly consider the forwardschein signal for extended sources of radiation.

\section{Sources}
\label{sec:source}
\subsection{Simple point sources}
The simplest class of sources of stimulating radiation are ones whose fluxes are stable on the light-crossing timescale of the inner Milky Way DM halo. In this limit, we can treat the source emission as constant in time and can simply integrate the gegenschein flux along a DM column without weighting by a varying brightness history. The brightest objects that fall into this category (and some of the brightest radio point sources observed at the present, like Cygnus A) are  quasars and radio galaxies, whose emission is powered by relativistic jets emerging from central supermassive black holes in galaxies. These point sources have the additional property that they are in the infinitely-far limit where the gegenschein geometry is independent of their distance and all gegenschein images have the same $\sim$arcminute spatial smearing, given by \eqref{eq:Ig_ps}.
\begin{figure}
    \centering
    \includegraphics[width=0.48\textwidth]{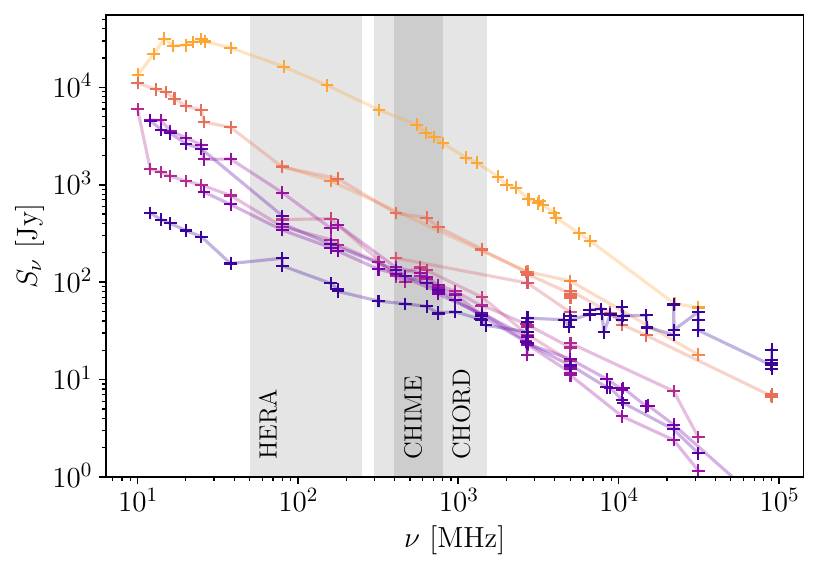}
    \caption{\textbf{Measured and interpolated spectra of the top 10 brightest extragalactic radio point sources at 1~GHz.} We consider the axion stimulated decay signal due to radio galaxies from the Keuhr catalog \cite{kuhr1981catalogue}. The black crosses and curves show the measured and linearly interpolated spectra of the sources, and the fluxes outside the observation range are assumed to be zero. The top yellow line corresponds to Cygnus A. The gray bands correspond to the frequency ranges of some telescopes we consider.}
    \label{fig:egrs_top_spec}
\end{figure}

For the contribution of stimulating radiation from extra-galactic radio sources, we refer to the Keuhr catalog \cite{kuhr1981catalogue} of 518 radio galaxies or quasars that are brighter than 1~Jy at 5~GHz. Since the Keuhr catalog is not complete within 10$^\circ$ of the galactic plane, we supplement the catalog with 22 additional extragalactic point sources brighter than 5~Jy at 1.4~GHz from the \texttt{cora} code package~\cite{cora2022}. Notably, this includes Cygnus A, which is by far the brightest radio point source in the sky. We additionally included spectral measurements of Cygnus A from \cite{baars1977absolute}, as its stimulating radiation is the dominant contribution out of all extragalactic point sources. \figref{fig:egrs_top_spec} shows the spectra of the top 10 brightest sources at 1~GHz. With the exception of the top spectrum (Cygnus A) all other spectra come from the Keuhr catalog. We linearly interpolate (in log-log space) flux measurements to construct a continuous spectrum and assume the fluxes outside of the measured range are zero. As discussed in Sec.~\ref{sec:sensitivity}, the contribution of simple point sources to stimulated axion decay is subdominant compared to other types of sources, so the final sensitivity does not depend sensitively on their assumed fluxes. 

\subsection{Short-duration transients}
Other radio point sources can be very bright for intermittent periods of time, including pulsars and fast radio bursts (FRBs). While the peak brightness of these sources can be high, the emission is generally too short-lived to contribute much to the signal. The gegenschein that would reach us at the present day is the culmination of axion decay stimulated along an entire DM column with different photon times of flight. Therefore, the relevant quantity is the \emph{average} brightness over the light-crossing time of the MW, which is generally quite low for pulsars and FRBs, rather than the peak brightness.  In other words, at a given observing time, a radio transient will only be able to stimulate decays over a very thin portion of the total DM column corresponding to the pulse or burst duration. This is in contrast to the stable point sources where the flux from all different column depths (corresponding to different photon times of flight) contributes to the signal. The enhanced brightness of radio transients on short timescales is not enough to compensate for the significantly shorter effective DM column that contributes to the gegenschein flux. Therefore, transient radio point sources can be neglected. The one exception is when the timescale for the source to dim appreciably is similar to the light crossing time of the DM halo, as is the case for SNRs described below.

\begin{figure*}
    \centering
    \includegraphics[width=0.99\textwidth]{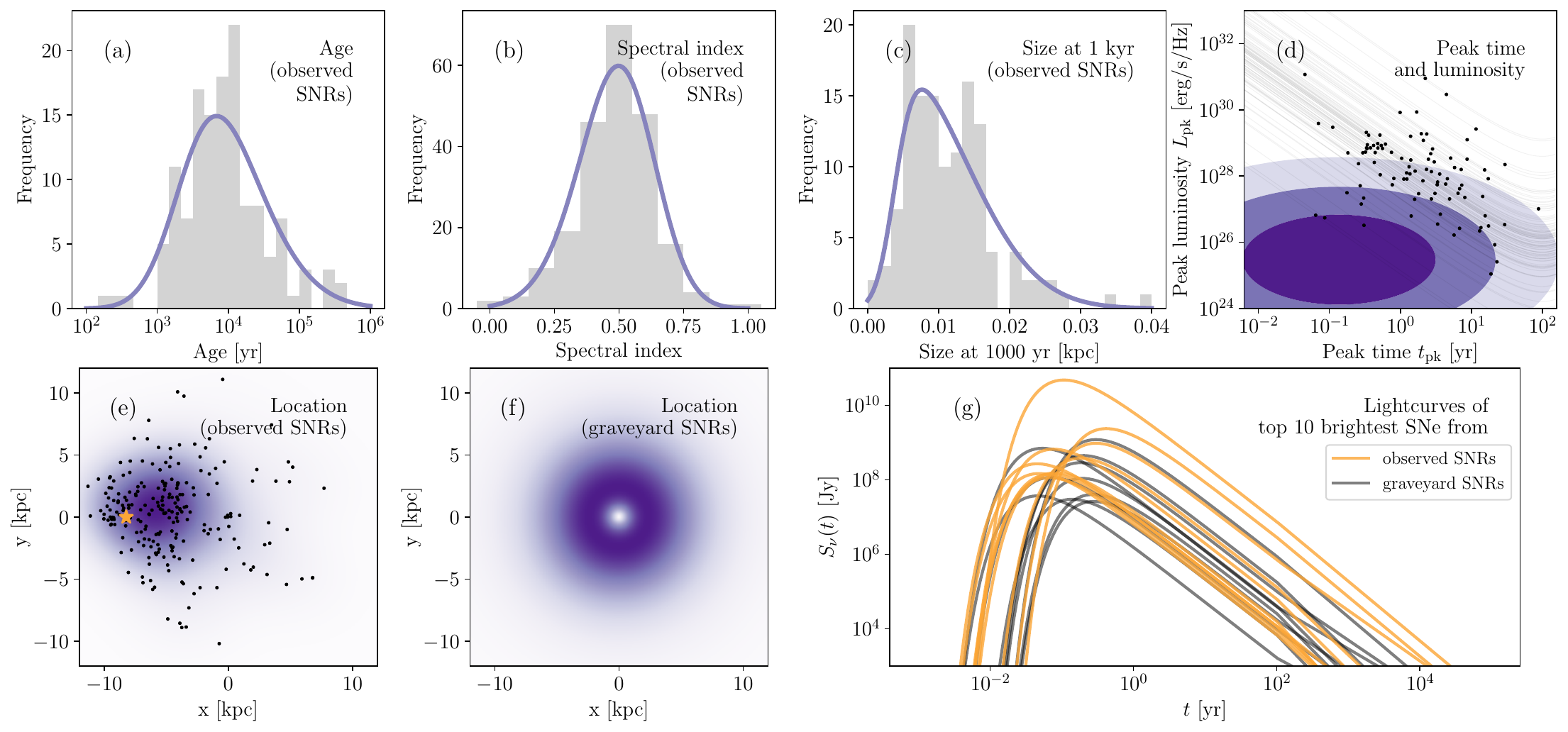}
    \vspace{-0.5cm}
    \caption{\textbf{Sampling missing information for observed and graveyard SNRs}
    Panels (a-c) show the distributions of age, spectral index, and physical size of the observed SNRs along with the best-fit skewed gaussian distributions.
    Panel (d) shows the 1-, 2-, and 3-sigma contours for the joint distribution of peak time and luminosity for a generic galactic SNR. For SNRs with observed fluxes, the grey lines show slices through this space that yield a predicted present-day flux that is consistent with measurement. For this population, our sampling is conditioned such that we draw from this 1D subspace for each SNR, with examples shown as black dots.
    Panel (e) shows observed SNR locations in a top-down view of our Galactic plane, along with the kernel density estimate of the distribution.
    Panel (f) shows the distribution of the graveyard SNRs.
    Panel (g) shows the lightcurves of the top 10 brightest SNRs in one out of 300 realizations.
    }
    \vspace{-0.3cm}
    \label{fig:snrpipeline}
\end{figure*}

\subsection{Supernova remnants} \label{sec:SNR}
SNRs can stay relatively bright for $\sim 10^4$ years, which is similar to the $\sim 10^4 - 10^5$~year light-crossing time of the inner MW halo. While SNRs can be very radio-bright now, they were substantially brighter in the past. Stimulating radiation from early phases of SNR evolution corresponds to axion decay in deeper parts of the DM column. The gegenschein signal strength benefits significantly from being able to integrate over the whole brightness history of the SNR by integrating over a given DM column.

In our analysis, we use known Galactic SNRs from Green's SNR catalog \cite{green2019revised} and SNRcat \cite{ferrand2012census}. We find 94 SNRs with measured ages and distances (along with both upper and lower bounds on age and distance), angular sizes, spectral indices, and fluxes. Additionally, there are 289 SNRs whose distances, ages, or spectral indices have not been fully characterized. On top of this less well-characterized population, as pointed out by Ref.~\cite{buen2022axion}, there is likely to be a large population of SNRs that are too dim to have even been detected at the present day due to their age or distance from Earth, but whose early phases of evolution can still contribute to the gegenschein signal. Following the nomenclature of Ref.~\cite{buen2022axion}, we refer to this population as the ``SNR graveyard.'' Since the focus of this work is to perform an all-sky forecast for observatories that survey a large fraction of the sky, we aim to include all relevant SNRs in a probabilistic way so that the aggregate signal strength is representative of the true one. In the discussion below, we detail how each relevant SNR parameter is either obtained from a dataset, modeled, or drawn from an empirical distribution. 

For the SNR graveyard, we assume Galactic supernovae occur as a Poisson process with a rate of 4.6 per century~\cite{adams2013observing} over the past 200,000 years, corresponding to an ``out-and-back'' light-travel distance of 30~kpc. We have performed convergence tests to ensure that even older SNRs do not contribute appreciably to the signal, since their early gegenschein images would lie outside of the inner Milky Way halo where the DM density is lower. We generate 300 random realizations with the unknown properties of the graveyard SNRs (e.g. distance, brightness history, etc.) drawn from empirical distributions described below. For observed SNRs with missing information, we similarly generate 300 random realizations. Within a given realization, we expect that errors on individual SNR properties (i.e. the difference between the true value and the value drawn from the empirical distribution) will wash out. Over such a large number of realizations, we expect that the spread in the resulting SNR signal strength should reflect the true distribution of the SNR contribution to the axion signal strength. We have checked that in 100 realizations, the 95\% coverage and median axion sensitivity are very similar, {differing by at most $\sim$10\% compared} to those obtained with 300 realizations, indicating convergence. Altogether, we have a large number of SNRs contributing to stimulating axion decay, including $\sim 400$ observed SNRs and $\sim9000$ unobserved SNRs; with 300 realizations of their properties, this corresponds to $\sim3$~million draws from the underlying distributions.

Modelling the SNR rate as a Poisson process results in a uniform distribution in time, with older SNRs contributing much less to the signal. However, as shown in panel (a) of \figref{fig:snrpipeline} the age distribution of the observed SNRs is very different due to observational biases, notably the fact that SNRs get dimmer as they age making them harder to detect. When the age information is missing for an observed SNR, or only a lower/upper bound is known, we sample from the empirically determined skewed gaussian age distribution shown in the same plot.

For SNRs in the graveyard, we follow Ref.~\cite{Green:2015isa} to determine their locations, taking an empirical surface density $\Sigma$ of the SNR graveyard in the Galactic plane, 
\begin{equation} \label{eq:spatialdist}
    \Sigma(R) \propto \left(\frac{R}{R_\odot}\right)^a \exp \left(- b \frac{R - R_\odot}{R_\odot}\right) 
\end{equation}
where $a = 1.09$ and $b = 3.87 $ based on a sample of 69 bright SNRs~\cite{Green:2015isa}. We assume the distribution in their height above the Galactic plane is exponential, so that the total 3D distribution follows
\begin{equation}
    p(\vx) \propto \Sigma(R) \exp(-\frac{|z|}{z_0}),
    \label{eq:height}
\end{equation}
where we take $z_0$ as the approximate scale height of the Galactic disk of 0.1~kpc \cite{adams2013observing}. As is the case for the SNR ages, the spatial distribution of the observed SNRs is different from the graveyard SNRs, and is heavily biased toward the Sun's location. Thus for observed SNRs with incomplete distance information, we perform a kernel density estimate of the SNRs with known distances $p_\text{obs}(\vx)$ as shown in \figref{fig:snrpipeline}. To sample the missing distance, we draw from the induced 1D distribution given the observed direction $\hat n$ of the SNR
\begin{equation}
    \tilde p(d)=p(\vx|\hat n~\dd\Omega)\propto p(d\hat n)\cdot d^2,
    \label{eq:kde}
\end{equation}
where the $d^2$ factor comes from the Jacobian $\dd^3\vx/\dd\Omega$. We take into account known lower or upper bounds on the distances as a prior when applicable.

To determine the brightness history of the SNRs, we follow the prescription of Ref.~\cite{sun2022axion} for modeling SNR light curves generated by synchrotron radiation of shock-accelerated electrons during the Sedov-Taylor phase. We additionally include the emission from the free expansion phase (before the Sedov-Taylor phase) in our analysis, taking an empirical approach based on observed lightcurves for young SNRs following Ref.~\cite{buen2022axion}. Notably, in contrast to these previous works, we are considering an observing strategy that includes a large ensemble of SNRs with properties drawn from a distribution over many realizations rather than focusing on a single SNR. Therefore, theoretical uncertainties in the brightness history from unobserved phases of the evolution (especially the free-expansion phase) are less likely to impact our resulting sensitivity projection. 

In the Sedov-Taylor phase, we take the specific synchrotron flux at distance $d$ to be given by the scaling $S_\nu \sim 1/d^2 V K_e B^{(p+1)/2} \nu^{-(p-1)/2} $ where $V $ is the volume where both electrons and the $B$ field are present, and where the electrons have a differential energy spectrum $d n_e /d \gamma = K_e \gamma^{-p}$ for Lorentz factor $\gamma$. The power law index of the energy distribution $p$ can be inferred from measurements of the SNR spectral index. For the SNR graveyard and for observed SNRs with unknown spectral indices, we draw from an empirically determined distribution of spectral indices using Green's catalog~\cite{green2019revised}. We find that a skewed-gaussian distribution is a good fit to the SNR spectral indices in the catalog, as shown in panel (b) of \figref{fig:snrpipeline}. We expect that this distribution is not significantly biased, since the spectral index does not play a substantial role in the observability of SNRs.

As our fiducial model, we assume that that the electron energy spectrum evolves according to the classic treatment of Ref.~\cite{shklovskii1960secular} where $V K_e \sim R^{1-p}$ for shock radius $R$. In Appendix~\ref{appd:model}, we include forecasts showing the predicted signal strength instead assuming an alternate model of adiabatic electron evolution~\cite{urovsevic2018foundation}, finding that this alternate modeling choice does not significantly impact the sensitivity. Further, we assume by default that the the magnetic field evolves as $B\sim R^{-2}$, which preserves the flux through the shock front. This is an intermediate scaling between that expected for resonant streaming instability $B$-field amplification, $B\sim R^{-1.5}$, and non-resonant amplification of the $B$ field, $B\sim R^{-2.25}$.
These different mechanisms can dominate the scaling at different parts of the Sedov-Taylor phase, so in Appendix~\ref{appd:model} we demonstrate that our forecasts are not sensitive to this modeling choice (similar to the case of the electron energy model). Finally, we determine the time evolution of the shock radius as $R \sim (E/\rho_{ISM})^{1/5} t^{2/5}$ during the Sedov Taylor phase. Note that, in contrast to our previous work~\cite{sun2022axion}, we include an empirical model of the free-expansion phase (described below) rather than conservatively assuming that radiation from before the Sedov-Taylor phase is generated by synchrotron radiation. Therefore, in the present analysis, we do not need to assume an onset time for the magnification of the SNR $B$-field, which was a key piece of information in our earlier work.

For the physical size of the SNRs in the graveyard, which affects the angular size of their axion gegenschein image, we again construct a skewed-gaussian distribution of the SNR size in Green's catalog, extrapolated to a reference time of 1000 years using the time-radius scaling relation in the Sedov-Taylor phase, and sample the SNR sizes from this distribution. This empirical distribution is shown in panel (c) of \figref{fig:snrpipeline}.

During the free-expansion phase, we adopt the empirical fitting form of Ref.~\cite{Bietenholz:2020yvw} which took a compilation of 1475 radio measurements of
294 young supernovae to determine a light curve,
\begin{equation}
    L(t) = L_\text{pk} e^{\frac{3}{2}(1 - t_\text{pk}/t) }\left(\frac{t}{t_\text{pk}}\right)^{-\frac{3}{2}} 
    \label{free_exp}
\end{equation}
where $L_\text{pk}$ and $t_\text{pk}$ are parameters that depend on the peak brightness of the SNR and that are drawn from a distribution. The best-fit distributions were found to be log-normal, with a mean $L_\text{pk}$ of $3 \times 10^{25}$~erg/s/ Hz with a standard deviation of 1.5 dex (taking into account the likelihood of many non-detections of SNRs at radio frequencies) and a mean $t_\text{pk}$ of 50 days with a standard deviation of 0.9 dex. We take $L_\text{pk}$ and $t_\text{pk}$ to be independent parameters (i.e. with no covariance). To match the free-expansion phase onto the Sedov-Taylor phase of evolution for observed SNRs with well-characterized properties, we draw free-expansion phase parameters from the empirical distribution, generate light curves, and compute the transition time $t_\text{free}$ when the luminosities as predicted by the free-expansion and Sedov-Taylor prescriptions match. The computed $t_\text{free}$ forms a distribution that peaks around $\sim$100 years after the SN, matching general expectations for the transition between the free-expansion and Sedov-Taylor phases. We fix this transition time to be 100 year for all SNR, since when $t\gg t_\text{pk}$, $L\sim t^{-1.5}$ in the free-expansion phase which is very similar to the power law behavior in the Sedov-Taylor phase ($L\sim t^{-1.6}$ for a SNR with the median spectra index of 0.5). In Appendix \ref{appd:model}, we show that changing this value to 30 or 300 years does not significantly change our sensitivity. To obtain the lightcurve for observed SNRs, i.e. their values of $L_\text{pk}$ and $t_\text{pk}$, we jointly sample the two variables, conditioned such that the lightcurve will lead to the observed flux $S_{\nu,\text{obs}}$ today
\begin{equation}
    p(L_\text{pk}, t_\text{pk}|S_{\nu,\text{obs}}) \sim p(L_\text{pk})p(t_\text{pk})\delta(S_\nu-S_{\nu,\text{obs}}).
\end{equation}
\begin{figure*}
    \centering
    \includegraphics[width=0.99\textwidth]{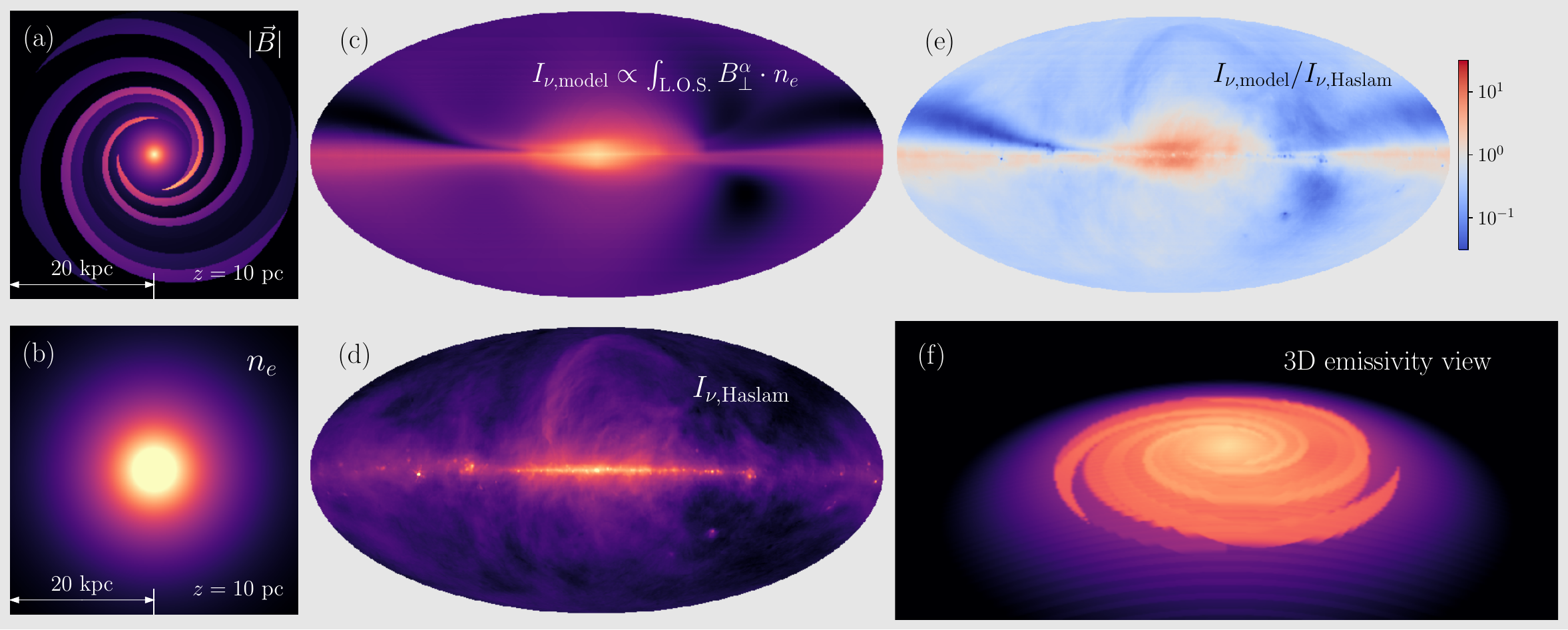}
    \vspace{-0.3cm}
    \caption{\textbf{Pipeline for estimating GSR emissivity.} In panels (a) and (b), we show our fiducial Galactic magnetic-field model from Ref.~\cite{jansson2012new} and the relativistic electron density assumed therein~\cite{cordes2002ne2001}. These models, along with the electron spectral index, allow us to determine the specific GSR intensity seen from Earth, shown in panel (c). We compare this intensity with the Haslam map of the radio sky shown in panel (d), and obtain the intensity ratio shown in panel (e), which captures small scale variations of the emissivity. Using this ratio together with the 3D model, we can construct the combined emissivity shown in panel (f).}
    \label{fig:gsrpipeline}
    \vspace{-0.4cm}
\end{figure*}

To summarize our sampling procedure for the SNR graveyard, we (1) draw their age from a uniform distribution assuming a constant SNR rate of 4.6 per century; (2) position them according to \eqref{eq:spatialdist} and \eqref{eq:height}; (3) determine the free-expansion phase lightcurve from \eqref{free_exp}, drawing $L_\text{pk}$ and $t_\text{pk}$ from empirically determined log-normal distributions, which then transitions into the Sedov-Taylor phase at 100 years; and (4) draw a spectral index and physical size from the empirically determined distribution based on Green's catalog. For observed SNRs (in the event of missing information), we (1) draw their age from an empirical distribution of observed ages; (2) determine their distance from an empirical kernel density estimate based on the observed locations of SNRs; (3) draw a spectral index; (4) jointly sample $L_\text{pk}$ and $t_\text{pk}$ such the predicted flux today matches the observed flux (if flux is measured) assuming $t_\text{free}=100$~yr and with the fiducial modeling assumptions during the Sedov-Taylor phase; and (5) draw the physical size from the empirically determined distribution based on Green's catalog. We discard any draws of SNR properties that yield an expected present-day flux greater than the brightest observed SNR at 400~MHz, and instead we re-draw from the distributions. This ensures that we are not ``double counting'' by invoking SNRs in the graveyard that would have been observed already. We have verified that doing so does not significantly change the total sensitivity obtained from the graveyard SNRs, which is sub-dominant to the contribution from observed SNRs.

In the sensitivity plot of Fig.~\ref{fig:reach}, the bands represent variations (95\% containment) coming from different realizations of the properties of observed and unobserved (graveyard) SNRs. In addition to the statistical uncertainties shown, we also consider the systematic uncertainties introduced by the discrete modeling choices we made for the SNR brightness history, finding this uncertainty to be subdominant as discussed in Appendix~\ref{appd:model}.

\subsection{Galactic synchrotron radiation}
\label{sec:gsr}
GSR is a diffuse, low surface brightness source. While it is not as bright as individual point sources, GSR covers a large fraction of the sky and is thus potentially an important contribution to the overall stimulated axion DM decay signal for all-sky searches. Synchrotron radiation is produced when high energy cosmic ray electrons interact with the galactic magnetic field, emitting in radio frequencies. Currently, the only all-sky observation of GSR comes from the de-sourced Haslam map at 408~MHz \cite{haslam1981408}, with a measured frequency scaling $\sim \nu^{-2.5}$ \cite{guzman2011all}. As shown in \figref{fig:gsrpipeline} (d), GSR is concentrated on the galactic plane and exhibits considerable variations on large angular scales, primarily due to the non-uniform production and propagation of cosmic ray electrons and the spatially varying structure of the galactic magnetic field. When compared to the resolution of radio telescopes like CHIME and the angular kernel from the DM velocity dispersion, the GSR is a smooth, diffuse source. The optical depth to the GSR is also much smaller than unity~\cite{schnitzeler2012modelling}, with the exception of the few arcmins around the Galactic center~\cite{1989ApJ...342..769P}, for which we discard the corresponding pixels in our analysis. The large-scale features of GSR emission are not likely to vary on timescales comparable to the light-crossing time of the MW.

In the frequency range we consider ($\nu \lesssim1$~GHz), GSR is the dominant foreground and background for axion stimulated decay, serving both as a source of (spectrally distinct) signal and noise. Therefore, unlike bright point sources, it is important to consider the forwardschein from GSR in addition to the gegenschein; for a point source, even one with strong time variation, the forwardschein has no time delay and the background for detecting the forwardschein is the bright point source itself, whereas for synchrotron radiation the forwardschein and gegenschein appear on equally ``blank'' parts of the sky. Taken together, despite the low surface brightness, the all-sky contribution of stimulated decay due to GSR is important due to its sheer spatial extent.

Unlike other sources we consider, the observed GSR is emitted along the entire line of sight (LOS). We therefore need to evaluate the full integral of \eqref{eq:Ist} over the source position. This integral includes both forwardschein and gegenschein in the same and opposite directions of the source through the kernel of \eqref{eq:Fie}. To perform this integral, we need to know where the GSR is emitted along the LOS, which requires us to consider 3D models of the galactic magnetic field and relativistic electron density. On the other hand, the empirical understanding we have of GSR comes primarily from the de-sourced Haslam map which only provides 2D information projected along the LOS. We therefore use a combination of the 2D observational information and 3D parametric models of the GSR emissivity to calculate the expected axion stimulated decay intensity. To do this, we choose a particular GSR model, i.e. a combination of a galactic magnetic field model and a relativistic electron distribution model, and keep only the coherent large-scale (regular) field. We compute the GSR emissivity as 
\begin{equation}
    j\propto B_\perp^{(p+1)/2}n_e
\end{equation}
where $B_\perp$ is the magnitude of the field perpendiculat to the LOS, and $p$ is the relativistic electron spectral index, which we assume to be $p=3$ \cite{tanabashi2018review, jansson2012new}. We then calculate the expected GSR emission observed at Earth's location, and then scale the normalization of the model's GSR emissivity in each direction independently such that the observed intensity matches the Haslam map at 408~MHz. By doing this, we have effectively used the Haslam map to restore the random variations of the transverse $B$ field that we did not explicitly include in our model, while using the 3D model to only inform the LOS distribution of the GSR emission. 

For our fiducial models, we use the non-random (regular) components of the $B$-field model in Ref.~\cite{jansson2012new}, shown in panel (a) of \figref{fig:gsrpipeline}, and the NE2001 model \cite{cordes2002ne2001} for relativistic electron distribution shown in panel (b) of the same figure, which is also used in Ref.~\cite{jansson2012new}. The calculated GSR intensity from these models are shown against the observed intensity (Haslam) map in panel (c) and (d). As expected, the Haslam map contains more random variations on small angular scales than the modelled map, though the two have a similar structure on large scales. Their ratio (adjusted such that the mean is 1) is shown in panel (e). Panel (f) shows an external view of the synchrotron emissivity (normalized by the Haslam map). Along with the DM density distribution, the GSR emissivity is used to calculate the total axion stimulated decay intensity.

In order to quantify the systematic errors from choosing this particular $B$-field model, we also consider two other older models of the galactic magnetic field in Ref.~\cite{sun2008radio}: the Axis-symmetry Spiral + Halo model and a Bi-symmetric Spiral + Halo model. In Appendix \ref{appd:model}, we show the $B$ field distribution of these alternative models and the expected stimulated decay sensitivity from considering GSR alone. The predicted sensitivities are very similar across different models, differing by around 1\% compared to the fiducial model. This makes the GSR a particularly robust contribution to the total axion sensitivity. This robustness can be attributed to the fact that the 3D models inform the LOS distribution of the GSR emissivity while the total normalization is primarily determined from the observed Haslam map. Even though GSR is a weaker contribution to stimulating radiation than SNRs, the sensitivity to axion decay induced by GSR is subject to much smaller uncertainties. Therefore, the sensitivity inferred from considering GSR only (omitting SNRs from the analysis) constitutes a minimal sensitivity to axions.

\section{Sensitivity} \label{sec:sensitivity}
For this analysis, we primarily consider compact radio interferometer arrays, with CHIME as our fiducial instrument. Such telescopes are optimized for mapping speed, such that they are well-suited to making deep maps of large portions of the sky. We also consider existing and planned compact arrays such as the Hydrogen Epoch of Reionization Array (HERA)~\cite{deboer2017hydrogen}, the Canadian Hydrogen Observatory and Radio-transient Detector (CHORD)~\cite{vanderlinde2019lrp}, the Hydrogen Intensity Real-time Analysis eXperiment~(HIRAX) \cite{crichton2022hydrogen}, and the Bustling Universe Radio Survey Telescope in Taiwan (BURSTT) \cite{lin2022burstt}.

Radio interferometers fundamentally measure modes on the Fourier plane of the sky, which can be mapped to the aperture plane of the telescope. However, since the interferometers considered here are compact, the aperture plane is filled at the order-unity level. We therefore make simplifying approximations in the following discussion that allow us to map the measurement into configuration space as if performed by a single-dish telescope. We then apply a correction for the missing Fourier modes due to the unfilled parts of the aperture.

In detail, the synthesized beam of an interferometer has a solid angle of $\Delta \Omega_\text{ideal} \sim \lambda^2 / A_\text{array}$, where $A_\text{array}$ is the extent of the area covered by the array. If the array area were completely filled, the brightness temperature of each element of the sky with size $\Delta \Omega$ and brightness temperature $T_\text{sig}$ would be measured with a signal-to-noise ratio 
\begin{equation} \label{eq:snratio}
    S/N = \frac{ T_\text{sig}}{T_\text{sky} + T_\text{receiver} / \eta_s} \sqrt{n_\text{pol}\Delta\nu~t_\text{obs}},
\end{equation}
where $T_\text{sky}$ is the brightness temperature of backgrounds on the sky, $\eta_s$ is the signal chain efficiency, $T_\text{receiver} / \eta_s$ is the sky-calibrated receiver noise temperature, (\textit{i.e.}~the receiver noise temperature accounting for any signal losses from the signal chain or correlator), and $n_\text{pol}$ is the number of polarizations observed by the telescope, which is 2 for all dish telescopes and 1 for BURSTT. Realistically, the full extent of the array is occupied by the physical receiver up to a filling factor
\begin{equation}
    \eta_f=A_\text{aperture} / A_\text{array},
\end{equation}
where $A_\text{aperture}$ is the physical area covered by the aperture. Aside from the issue of not spatially filling the full array area, the aperture is not completely efficient in capturing the incoming power, leading to another fractional loss of power characterized by the aperture efficiency
\begin{equation}
    \eta_a=A_\text{eff} / A_\text{aperture},
\end{equation}
where $A_\text{eff}$ is the effective aperture area after accounting for losses. Due to the unfilled array area, the synthesized beam obtains sidelobes such that measured pixels on the sky are no longer independent. A simple way to approximately account for this is through the use of an \emph{effective} beam solid angle,
\begin{equation} \label{eq:deltaomega}
    \Delta \Omega = \lambda^2 / A_\text{eff} = \Delta\Omega_\text{ideal}/(\eta_f\,\eta_a)
\end{equation}
such that the measured brightness within each effective beam constitutes a measurement that is approximately independent of the other effective beams. In our sensitivity estimates, we will use Cartesian pixelizations of the sky with pixel sizes consistent with the above expression, which will allow us to treat each pixel as an independent measurement. Note that only $A_\text{eff}$ determines the density of pixels or number of independent measurements in any direction on the sky, which will be the driving factor in determining the sensitivity. The exact geometry of the array, as long as it is compact, only affects the relative locations of the pixels, and is secondary in determining the overall sensitivity. We note additionally that this treatment is most accurate for arrays with a maximally redundant baseline configuration, \emph{e.g.} square or hexagonal arrays, like all the arrays we consider in this work. For arrays with less redundant baseline configurations, such as the Murchison Widefield Array (MWA) and Low-Frequency Array (LOFAR), the beams remain independent but their sensitivity is reduced. Since they do not repeatedly sample the same redundant baselines for measuring the spatially extended axion signal, the sensitivity using non-redundant baselines scales as $\sqrt{A_\text{array} / A_\text{eff}}$ relative to the redundant arrays.

The telescopes considered here are drift-scan telescopes, in that during observations they have fixed pointing and scan the sky only via the Earth's rotation. For any location on the sky, the integration time per day is given by the time it stays in the primary beam of the telescope
\begin{equation} \label{eq:tobs}
    t_\text{obs}/N_\text{days} = 24\text{~hour}\cdot \frac{\Delta\theta_\text{primary,EW}}{2\pi \cos\delta},
\end{equation}
where $\Delta\theta_\text{primary,EW}$ is the angular extent of the primary beam in the east-west (EW) direction. We note that the primary beam is associated with one dish in an array, and differs from the synthesized beams formed by multiple dishes. The $\cos\delta$ factor (where $\delta$ denotes the declination) accounts for the fact that a LOS closer to the North or South Poles will have a longer integration time (assuming a fixed EW angular extent of the primary beam). If the poles are in the telescope field of view (FOV), we cap the integration time for pixels near the poles at 24 hours per day. Additionally, for telescopes with a FOV on both sides of either pole, pixels around the pole would enter the instantaneous FOV twice a day, in which case we need to double the integration time in \eqref{eq:tobs}. In the telescopes we consider, only CHIME observes any polar region of the sky. To compare the axion sensitivities between different instruments, we assume $t_\text{obs}=5$~years of observation time for each one, matching the existing collecting time of CHIME (corresponding to archival data). Alternatively, when only accounting for data taken at night, 5 years roughly corresponds to the ultimate expected exposure of all CHIME data. Our sensitivity forecasts depend only weakly on this assumption, with the reach to $\gagg$ scaling as $\sim t_\text{obs}^{1/4}$.

The signal we expect to see is a spectral line from stimulated decay that is broadened by the DM velocity dispersion, which we approximate as a Gaussian with width $\sigma_d$. Following Refs.~\cite{ghosh2020axion,sun2022axion}, we take a top hat spectral window of width $\Delta\nu=2.17\nu_d\sigma_d$ centered at the decay frequency $\nu=m_a/4\pi$ to maximize the signal-to-noise ratio. This admits a fraction $f_\Delta=0.721$ of the total signal power, which we take into account in all signal temperature calculations. Since $\sigma_d\sim10^{-3}$, achieving the top hat window function requires the telescope to have a resolving power $\nu/\Delta\nu$ greater than 1000, which is satisfied by all the telescopes we consider. We expect that some frequencies will be contaminated by human-generated radio frequency interference (RFI). If the RFI is persistent at certain frequencies, then the search will lose sensitivity to the corresponding axion masses. For intermittent RFI, the resulting in a loss of integration time is unlikely to substantially affect our forecasts due to the weak scaling of our sensitivity to $\gagg$ with $t_\text{obs.}$ as discussed above.

In general, we will use a grid pixelization in right ascension and declination. The pixelization and integration time for each telescope we consider differ slightly, as described in detail in the following Subsection. In the Subsection after that, we then summarize our estimate procedures and attempt to derive a figure of merit for the ability of the telescopes we consider to detect axion DM.

\subsection{Telescopes}
\subsubsection{CHIME}
Our fiducial instrument, CHIME, is a stationary transit telescope that observes in the frequency range of 400-800~MHz. It consists of four 100~m$~\times~20~$m cylindrical reflectors whose cylinder axes are aligned along the north-south (NS) direction. There are 2~m gaps between neighboring reflectors, making the EW extent $D_\text{EW} = 86$~m. In the EW direction, we therefore pick up a filling factor $\eta_{f,\text{EW}}=80~\text{m}/86~\text{m}\approx0.93$. The aperture illumination of CHIME is very uniform in the NS direction, with a illuminated cylindrical length of $l_\text{cyl}=78~$m.  Situated at $49.3^\circ$N, CHIME can see in declination from approximately 10$^\circ$S to 19$^\circ$ past the North Pole, with directions from $71^\circ$~N to $90^\circ$~N entering the instantaneous FOV twice a day \cite{amiri2022overview}.

To build a rectilinear pixelization in equatorial coordinates, we construct a grid that is uniform in $\alpha$ (right ascension) and varying in $\delta$ (declination) such that \eqref{eq:deltaomega} holds consistently for $\delta$-dependent $A_\text{eff}$. The extent of the pixels in the declination direction can be expressed as
\begin{equation}\label{eq:CHIMEdeltaDEC}
    \Delta \delta = \lambda / (D_\text{NS}\cos\delta_Z\cos\delta),
\end{equation}
where $D_\text{NS} = l_\text{cyl} = 78$\,m, and $\delta_Z=\delta-\delta_\text{CHIME}$ is the zenith angle, \emph{i.e.} the difference between declination of the pixel and the latitude of CHIME. The $1/\cos\delta_Z$ factor accounts for the fact that CHIME has a smaller baseline for LOS directions that are not directly above the instrument. The $1/\cos\delta$ factor compensates the reduced pixel width in right ascension as declination increases, common to all telescopes. This factor of $1/\cos\delta$ would ideally be incorporated in the right ascension pixel width $\Delta\alpha$. However, to keep a simpler rectilinear pixelization, we have shifted this factor to the declination pixel width $\Delta\delta$. We have checked that the polar regions of the sky (where this shift is most pronounced) do not significantly contribute to our sensitivity, since our signal is dominated by SNRs near the Galactic plane. Therefore, we do not expect that changing these details of our pixelization procedure will impact the resulting sensitivity.

In the EW direction, the short focal-length cylinders of CHIME are poorly illuminated. As such, the pixelization in right ascension $\alpha$ can be expressed as
\begin{equation}
    \Delta\alpha = \lambda / (\eta_{f,\text{EW}} \eta_a D_\text{EW})
\end{equation}
where $\eta_{f,\text{EW}}D_\text{EW} = 80$~m as noted above. The EW extent of the primary beam can be expressed as
\begin{equation}
    \Delta\theta_\text{primary,EW} = \lambda / (\eta_a D_\text{EW,primary}),
\end{equation}
where $D_\text{EW,primary} = 20$~m is the width of a single reflector. Using Fig. 17 of Ref.~\cite{amiri2022overview}, which shows measurements of full width half max (FWHM) $\Delta\theta_\text{EW,primary}$, we therefore deduce that $\eta_a\approx0.5$. Note that we have attributed the effective loss of sensitivity entirely to the EW direction, since only the EW direction contains gaps and is poorly illuminated. Based on the pixelization and efficiencies described here, this corresponds to 17.2 minutes on the equator for 400~MHz and 8.6 minutes for 800~MHz.

To estimate the receiver temperature of CHIME, we determine the calibrated temperature on a dim part of the sky (shown as the blue curves in Fig. 25 of Ref.~\cite{amiri2022overview}) and subtract off the expected synchrotron radiation temperature in the corresponding location in the CHIME FOV. We then average the remainder over frequency in order to estimate the calibrated receiver temperature, which is approximately 40~K.

\subsubsection{HERA}
HERA is a stationary transit telescope situated in South Africa at $30.7^\circ$~S, observing at 50-250~MHz \cite{deboer2017hydrogen}. Currently under construction and taking data with the existing elements, the full HERA instrument will feature 350 of $D_\text{dish}=14$~m diameter dishes, $N_\text{core}=320$ of which will be arranged in a hexagonal pattern in the compact core. Because of the hexagonal arrangement, the observed fourier modes cannot be simply translated to a pixelization on the sky. Instead, we will continue to use a pixelization that is rectilinear in equatorial coordinates. This approximation is equivalent to considering HERA to be a square array of the same total array size with the same geometric filling factor $\eta_f$.

Focusing on the core elements, the dishes are arranged in a hexagonal pattern with the dish centers 14.6~m apart. Each hexagon occupies $A_\text{hex}=\frac{\sqrt{3}}{2}(14.6~\text{m})^2\approx184.6~\text{m}^2$. The circular dish occupies $A_\text{dish}=\frac{\pi}{4}(14~\text{m})^2\approx153.9~\text{m}^2$, making the filling factor $\eta_f=A_\text{dish}/A_\text{hex}\approx0.834$. We will approximate this array as a square with side $D=\sqrt{N_\text{core}A_\text{hex}}\approx 243~\text{m}$, and the same filling factor $\eta_f$. We thus have the following pixelization
\begin{equation} \label{eq:HERAdeltaDEC}
    \Delta\delta=\lambda/(\sqrt{\eta_f\eta_a}D\cos\delta_Z\cos\delta),
\end{equation}
and
\begin{equation} \label{eq:HERAdeltaRA}
    \Delta\alpha=\lambda/(\sqrt{\eta_f\eta_a}D),
\end{equation}
where the factors $1/\cos\delta_Z$ and $1/\cos\delta$ have the same origin as those in \eqref{eq:CHIMEdeltaDEC} (with $\delta_Z=\delta-\delta_\text{HERA}$). Note that we have distributed the pixel size increase due to $\eta_f$ and $\eta_a$ evenly in the declination and right ascension directions due to the symmetry of EW and NS directions under our square array approximation. We take the aperture efficiency of HERA to be $\eta_a=0.6$~\cite{deboer2017hydrogen}.

We approximate the primary beam FWHM of the HERA dishes as
\begin{equation}
    \Delta\theta_\text{primary}=\frac{\lambda}{\sqrt{\eta_a}D_\text{dish}},
\end{equation}
where $D_\text{dish}=14$~m. This gives a FWHM of $11.6^\circ$ at 137~MHz, which is roughly consistent with the measured FWHM of $\sim10^\circ$~\cite{deboer2017hydrogen}. The primary beam size determines both the instantaneous FOV in the NS direction and the integration time for a fixed location on the sky, as the telescope transits in the EW direction. Notably, the instantaneous FOV is much smaller than that of CHIME. The integration time can be determined from \eqref{eq:tobs}, which corresponds to 147-29 minutes in the center of the FOV, at 50-250~MHz respectively. Finally for the calibrated receiver temperature of HERA, we take $T_\text{receiver} / \eta_s=100$~K.

\subsubsection{CHORD}
CHORD is a partially constructed radio telescope array situated at 49.3$^\circ$N. It is proposed to feature a 512-dish compact core of ultra-wideband dishes covering 300-1500~MHz \cite{vanderlinde2019lrp}. While the outrigger stations feature additional telescopes, we will focus on the core array, which is laid out in a rectangular grid. We approximate the array as 22 grid points in the EW direction and 23 in the NS direction, with each grid space occupying a 7~m by 9~m rectangle (7~m in the EW direction), making the total array extent $D_\text{EW}\times D_\text{NS}=154~\text{m}\times 207~\text{m}$. CHORD's dishes have diameter $D_\text{dish}=6$~m, making the filling factor $\eta_f=\frac{\pi}{4}(6~\text{m})^2/63~\text{m}^2\approx0.45$. We estimate the aperture efficiency to be $\eta_a=0.5$. Although the geometry is not symmetric between the EW direction and the NS direction, due to complicated shape of the gap, we approximate the fourier mode loss in both directions as the same.

CHORD differs from CHIME (and HERA) in that its dishes can be manually repointed in elevation between observing campaigns, meaning that although the instantaneous FOV is narrow, the total survey area can cover a large angle in the NS direction up to 30$^\circ$ on either side of the zenith. This also implies that the total effective area will not necessarily suffer the reduction due to zenith angle, a factor of $\cos\delta_Z=\cos(\delta-\delta_\text{CHORD})$. Our pixelization will be
\begin{equation} \label{eq:CHORDdeltaDEC}
    \Delta\delta=\lambda/(\sqrt{\eta_f\eta_a}D_\text{NS}\cos\delta)
\end{equation}
and
\begin{equation} \label{eq:CHORDdeltaRA}
    \Delta\alpha=\lambda/(\sqrt{\eta_f\eta_a}D_\text{EW}).
\end{equation}
We similarly approximate the primary beam FWHM as
\begin{equation} \label{eq:CHORDthetaprimary}
    \Delta\theta_\text{primary}=\lambda/(\sqrt{\eta_a}D_\text{dish}),
\end{equation}
where $D_\text{dish}=6$~m. Since CHORD is a pointing telescope array, the exact integration time of each pixel depends on the survey strategy. As an approximation, we assume CHORD is going to uniformly cover its full survey declination, from $19^\circ$N to $79^\circ$N, spanning $\Delta\theta_\text{survey}=60^\circ$. The integration time for each pixel will be modified from \eqref{eq:tobs} by the survey time fraction
\begin{equation} \label{eq:CHORDtobs}
    t_\text{obs}/N_\text{days} = 24\text{~hour}\frac{\Delta\theta_\text{primary,EW}}{2\pi \cos\delta}\frac{\Delta\theta_\text{primary,NS}}{\Delta\theta_\text{survey}},
\end{equation}
where $N_\text{days}$ correspond to the total number of days in the full survey. We take the calibrated receiver temperature $T_\text{receiver}/\eta_s=30$~K \cite{vanderlinde2019lrp}.

\subsubsection{HIRAX}
HIRAX is an array of dish telescopes of similar layout to CHORD currently under development. The initial HIRAX-256 features a $16\times16$ array of 6~m dishes observing at 400-800~MHz, with future plans to expand into a $32\times32$-element array (HIRAX-1024). Situated at $30.7^\circ$~S, it is expected to observe up to a zenith angle of $30^\circ$, from $0^\circ$-$60^\circ$~S thanks to its pointing dishes~\cite{crichton2022hydrogen}, similar to those of CHORD. The grid configuration of HIRAX will also be similar to CHORD, with each dish occupying a $7~\text{m}\times9~\text{m}$ area. As noted below \eqref{eq:deltaomega}, the detailed arrangement of the telescope dishes does not significantly affect our sensitivity projection. As with CHORD, we similarly assume an aperture efficiency of $\eta_a=0.5$. We use \eqref{eq:CHORDdeltaDEC} and \eqref{eq:CHORDdeltaRA} for the sky pixelization of HIRAX, and additionally use the expressions for the primary beam width and integration time in \eqref{eq:CHORDthetaprimary} and \eqref{eq:CHORDtobs}, with the appropriate factors substituted in for HIRAX instead of CHORD. We take the calibrated receiver temperature of HIRAX to be $T_\text{receiver}/\eta_s=50$~K \cite{crichton2022hydrogen}.

\subsubsection{BURSTT} 
BURSTT will be a compact antenna array situated at approximately $23.7^\circ$~N, observing in 300-800~MHz \cite{lin2022burstt}. BURSTT-256 will feature a main station of $16\times16$ array of antennas, each occupying a $2~\text{m} \times 2~\text{m}$ area, with plans to extend to a 2048 element array. In contrast to the dish arrays considered above, the BURSTT antennae yield a very large instantaneous FOV in both the NS and EW directions. The antennae also feature a relatively consistent beam width at $\Delta\theta_\text{primary}\approx 60^\circ$ in both NS and EW directions, throughout the observing frequency range, making the effective collecting area frequency dependent. This can be expressed as
\begin{equation}
    \Delta\theta_\text{primary}=\lambda\Big/\sqrt{A_\text{ant,eff}(\lambda)}=\lambda_0/\sqrt{A_{\text{ant,eff},0}},
\end{equation}
where $\lambda_0$ is a reference wavelength and $A_{\text{ant,eff},0}$ is the corresponding effective area of a single antenna. We thus have
\begin{equation}
    A_\text{eff}(\lambda)=A_{\text{eff},0}~\lambda^2/\lambda_0^2.
\end{equation}
Choosing $\lambda_0=c/300$\,MHz, the effective area is $A_{\text{eff},0}=0.91~\text{m}^2$, making the collecting efficiency
\begin{equation}
    \eta(\lambda)=A_\text{eff}(\lambda)/A_\text{array}
\end{equation}
approximately 0.23 at the reference frequency of 300~MHz. Note that $\eta$ corresponds to the product of the filling factor and aperture efficiency $\eta_f\eta_a$ for dish arrays. The pixelization of BURSTT can be determined similarly to the other telescopes we consider using \eqref{eq:HERAdeltaDEC} and \eqref{eq:HERAdeltaRA}, where we substitute $\eta_f\eta_a$ with $\eta(\lambda)$, obtaining
\begin{equation}
    \Delta\delta=\frac{\lambda}{\sqrt{\eta(\lambda)A_\text{array}}\cos\delta_Z\cos\delta} =\frac{\Delta\theta_\text{primary}}{\sqrt{N}\cos\delta_Z\cos\delta},
\end{equation}
and
\begin{equation}
    \Delta\alpha=\frac{\lambda}{\sqrt{\eta(\lambda)A_\text{array}}}=\frac{\Delta\theta_\text{primary}}{\sqrt{N}},
\end{equation}
where $N=256$ is the number of elements in the array. The integration time for each pixel can be calculated from \eqref{eq:tobs}. Note that in contrast to traditional dish telescopes, BURSTT will only observe one polarization. Finally, we take the calibrated receiver temperature to be $T_\text{receiver}/\eta_s=30$~K for BURSTT~\cite{lin2022burstt}.

\subsection{Figure of merit} \label{sec:merit}
\begin{table*}[ht]
    \centering
    \begin{tabular}{|C{2.5cm}|C{1.5cm}|C{1.2cm}|C{1.2cm}|C{2.5cm}|C{1.2cm}|}
        \hline
        Telescope arrays & Frequency [MHz] & $A_\text{eff}$ [$\text{m}^2$] & $\Omega_\text{FOV}$ [$\text{deg}^2$] & Relative merit $\sqrt{n_\text{pol}A_\text{eff}\Omega_\text{FOV}}$ & Relative $S/N$ \\
        \hline
        CHIME          & 400 & 3120  & 515  & 1.000 & 1.000 \\
        HERA           & 250 & 29548 & 40   & -     & -     \\
        HERA (extrap.) & 400 & 29548 & 16   & 0.537 & 0.537 \\
        CHORD          & 400 & 7173  & 102  & 0.676 & 0.993 \\
        HIRAX-256      & 400 & 3629  & 102  & 0.481 & 0.684 \\
        HIRAX-1024     & 400 & 14515 & 102  & 0.962 & 1.834 \\
        BURSTT-256     & 400 & 131   & 3600 & 0.383 & 0.251 \\
        BURSTT-2048    & 400 & 1037  & 3600 & 1.078 & 1.258 \\
        \hline
    \end{tabular}
    \caption{\textbf{Figures of merit and sensitivity comparison.} The figure of merit of \eqref{eq:fom} (normalized relative to CHIME) provides an estimate of how suitable an array is for probing axion stimulated decay. We also show the signal-to-noise ratio normalized to that of CHIME as computed with our full pipeline. In general, telescopes with large effective collecting area and instantaneous FOV are best suited for the axion search considered in this work. We compare different telescopes at 400~MHz, with the exception being HERA whose top frequency is 250~MHz. We use the frequency scaling in \eqref{eq:fom} to extrapolate the figure of merit for a comparison. At 250~MHz, the calibrated receiver temperature of HERA is comparable to the background sky temperature, so we use the full frequency scaling of the $T_\text{sig}/(T_\text{sky}+T_\text{rec}/\eta_s)$ factor to extrapolate the full $S/N$ we computed at 250~MHz up to 400~MHz for comparison. The extrapolation up to 400~MHz in the figure of merit and $S/N$ for HERA agree exceptionally well, highlighting the effectiveness and self-consistency of the various approximations used in deriving these scalings.}
    \label{table:merits}
    \vspace{-0.2cm}
\end{table*}

To intuitively understand the signal-to-noise performances of the various telescope arrays, and to guide future search efforts, in this Subsection we derive a figure of merit for telescopes under simplifying assumptions. We start from the total signal-to-noise ratio written as the sum in quadrature of the per-pixel ratio  \eqref{eq:snratio}
\begin{equation} \label{eq:SNperpixel}
(S/N)^2=\sum_i\left(\frac{T_{\text{sig},i}}{T_{\text{sky},i}+T_\text{rec}/\eta_s}\right)^2n_\text{pol}\,\Delta\nu\,t_{\text{obs},i},
\end{equation}
where $i$ indexes over the pixels. The temperature dependence in the right hand side is determined by the survey region of the telescope and relative brightness of the sky compared to the calibrated receiver temperature. In general, since our sensitivity is dominated by SNRs concentrated on the galactic plane, with higher concentration on the galactic center, we expect telescopes capable of seeing the galactic center or its antipodal point to have better sensitivity. The average GSR temperature measured in the Haslam map at 408~MHz is about 35~K, which is similar to the calibrated receiver temperature of a typical radio telescope we consider. Since the GSR flux scales steeply with the frequency as $\nu^{-2.5}$, we expect the receiver temperature to be the dominant source of systematic temperature only at high frequencies. At low frequencies, the temperature dependence therefore depends mainly on what parts of the sky are being surveyed.
In fact, since $T_\text{sig}$ is the highest near the galactic plane, the relevant $T_\text{sky}$ entering \eqref{eq:snratio} is also considerably higher, making the calibrated receiver temperature less dominant in its contribution to the system temperature at frequencies just above 400~MHz (the lower end of the frequency range covered by CHIME).

To obtain a simple figure of merit for comparing telescopes searching for stimulated axion decay, we neglect the temperature dependence (since it depends mainly on the properties of the sky at low frequencies rather than on the telescope) and focus instead on the remainder of the expression
\begin{equation}
    (S/N)^2\propto n_\text{pol}\,\nu\sum_i t_{\text{obs},i},
\end{equation}
where we have used the fact that the selected frequency band $\Delta\nu\propto\nu$.
Since the size of the instantaneous FOV does not change as the telescope drifts across the sky, the total integration time per day summed over all the pixels is simply 24 hours times the number of pixels in the instantaneous FOV $N_\text{pix,FOV}$,
\begin{equation}
    \sum_i t_{\text{obs},i}=24~\text{hour}\cdot N_\text{pix,FOV}.
\end{equation}
Therefore,
\begin{equation}
\begin{aligned}
    (S/N)^2&\propto n_\text{pol}\,\nu\, N_\text{pix,FOV}=n_\text{pol}\,\nu\,\frac{\Omega_\text{FOV}}{\Delta\Omega} \\
    &\propto \nu^3\,A_\text{eff}\Omega_\text{FOV},
\end{aligned}
\end{equation}
where in the last step we used \eqref{eq:deltaomega}. Thus, a figure of merit for the telescopes can be written as
\begin{equation}
    S/N \propto \nu^{3/2}\sqrt{n_\text{pol}A_\text{eff}\Omega_\text{FOV}},
\end{equation}
which is closely related to the \'etendue of the telescope $E=A_\text{array}\Omega_\text{FOV}$. For compact circular dish arrays and antenna arrays, the FOV is directly determined by the per-element effective area,
\begin{equation}
    \Omega_\text{FOV}\sim\frac{\lambda^2}{A_\text{eff,element}},
\end{equation}
which can further simply the figure of merit to
\begin{equation}
\begin{aligned}
    S/N &\propto \nu^{1/2}\sqrt{n_\text{pol}A_\text{eff}/A_\text{eff,element}} \\
    &= \nu^{1/2} \sqrt{n_\text{pol}N_\text{element}}. \label{eq:fom}
\end{aligned}
\end{equation}
The number under the square root can be roughly interpreted as the total number of channels at which the array is observing.

In Tab.~\ref{table:merits}, we compare the above figures of merit of the various telescope arrays we consider, which roughly aligns with the projected signal-to-noise ratio in a full numerical evaluation. In general, arrays with a large effective collecting area and instantaneous FOV are suitable for our search, since the expected signal comes from all parts of the sky (especially the Galactic plane). We compare all telescopes at a reference frequency of 400~MHz, except HERA which has a maximum frequency of 250~MHz. To compare HERA to the other instruments, we extrapolate its properties from 250~MHz to higher frequencies. For the figure of merit, this amounts to a simple frequency scaling in $\Omega_\text{FOV}$.

To understand the frequency scaling of the full sensitivity to axions, we must factor in the frequency dependence of the temperature-dependent factor in \eqref{eq:SNperpixel}. The signal temperature $T_\text{sig}\propto I_{g}(\nu)/\nu^2\Delta\nu$ depends on the spectral index of the dominant contribution to stimulated decay (SNRs) and how resolved the gegenschein images are by the interferometers. Taking the median SNR spectral index (panel b of \figref{fig:snrpipeline}) and assuming the SNR gegenschein image is resolved so that the solid angle size is fixed, then $I_{g}\propto\nu^{-0.5}$ and $T_\text{sig}\propto\nu^{-3.5}$. Meanwhile, the noise temperature scales differently at different frequency ranges: At high frequencies $T_\text{rec}/\eta_s \sim \nu^0$ dominates over $T_\text{sky}$, and at low frequencies, $T_\text{sky}\propto\nu^{-2.5}$ dominates. Put together, $S/N\propto\nu^{-3}\text{ or }\nu^{-0.5}$ at high and low frequencies, respectively. The sensitivity to axions scales as $\gagg\propto(S/N)^{-1/2}\propto\nu^{1.5}\text{ or }\nu^{0.25}$ at high and low frequencies, respectively. Although the estimate presented here involves many assumptions, this scaling argument agrees fairly well with the sensitivity projected in \figref{fig:reach}, which was computed fully generally without making such assumptions.

\subsection{Results}
In Fig.~\ref{fig:reach} we summarize the sensitivity to stimulated axion DM decay using existing or near-future radio telescopes, along with other constraints that are relevant to the accessible parameter space. There is clear complementarity between stimulated decay searches and the existing constraints from terrestrial experiment and observations of high-energy astrophysical phenomena.

Without any need for new hardware, CHIME, CHORD, HIRAX, and BURSTT can all constrain axion DM in mass gaps that are currently not constrained by haloscope experiments searching for axion conversion in a resonant cavity~\cite{panfilis1987limits, wuensch1989results, hagmann1990results, Asztalos:2003px, Boutan:2018uoc, ADMX:2018gho, ADMX:2021nhd, ADMX:2019uok, HAYSTAC:2018rwy, HAYSTAC:2020kwv}. CHORD and BURSTT also have sensitivity to slightly lower masses beyond what terrestrial experiments can currently access. No part of the axion DM parameter space accessible to HERA has been probed by terrestrial axion experiments to date.  
 
Meanwhile, other astrophysical probes can set powerful constraints on the existence of axions, irrespective of axions being the DM of our Universe. For instance, CAST sets very robust limits on axions emitted from the Sun which would be converted to X-ray photons in a laboratory magnetic field~\cite{Anastassopoulos:2017ftl}. Additionally, axions could be produced in magnetic white dwarfs and convert to X-rays in the magnetosphere~\cite{Dessert:2021bkv} or could induce a linear polarization in thermal magnetic white dwarf emission~\cite{Dessert:2022yqq}. Based on our forecast, the search for stimulated axion decay should yield an improved sensitivity compared to these searches (subject to our assumption that axions are all of the DM). The strongest astrophysical constraint in the parameter space of interest comes from pulsar polar caps where the plasma cannot screen electric fields, resulting in the emission of axions that can subsequently convert to photons resonantly~\cite{Prabhu:2021zve}; requiring that this emission not exceed the observed flux results in a strong constraint, shown as a line in Fig.~\ref{fig:reach} with the corresponding systematic uncertainty shown as a shaded band~\cite{Noordhuis:2022ljw}. Our median projected sensitivity is slightly weaker than the constraint from pulsar polar caps, but the 95\% containment for HERA does overlap with the pulsar polar cap systematic uncertainty band. Even absent the ability to access different parameter space, the search strategy presented here will still be a valuable cross check on searches involving high-energy astrophysical phenomena, since the systematic uncertainties related to those phenomena are completely orthogonal to those relevant to stimulated axion DM decay.

\section{Discussion and Outlook}
\label{sec:conclusions}
In this work, we have analyzed the sensitivity of a variety of survey interferometers to the stimulated decay of axion DM. Because of the wide FOV of these telescopes, we have carried out a systematic study of all possible astrophysical sources of stimulating radiation over the last $\sim 10^5$ years (corresponding to the Galactic light-crossing timescale). 

We find that extragalactic point sources and short-duration transients contribute negligibly to the overall signal. We additionally find that the strength of the stimulated decay signal from GSR is robust to different choices for modelling Galactic magnetic fields and relativistic electron densities. Therefore, the signal induced by GSR is subject to relatively small systematic uncertainty and constitutes the minimum sensitivity of these searches. 

On top of this minimum sensitivity, we find that stimulating radiation from SNRs can lead to an even larger axion decay signal. However, the SNR contribution to the sensitivity is subject to larger systematic uncertainties because the SNR emission history varies on timescales that are comparable to the Galactic light-crossing time. To determine the signal strength, we must therefore integrate over the whole SNR emission history, including parts that are either unmeasured or difficult to model. For instance, even for well-characterized SNRs, it is difficult to model the luminosity in the free-expansion phase when the SNR is brightest. Moreover, there may be a population of as-yet undetected SNRs (the SNR graveyard) that are dim today but whose stimulating radiation can still contribute to the signal. To quantify and mitigate these systematic uncertainties, we generated 300 realizations of signal templates based on drawing any unknown SNR properties from empirical distributions. For instance, we determined the lightcurves in the free-expansion phase using radio observations of young SNRs. Although the brightness of individual SNRs (and therefore, their contribution to the signal) can vary substantially, we find that in the aggregate, the variations wash out; the all-sky sensitivity to $\gagg$ varies by a factor of $\sim 2$ between median realizations and realizations at the edge of our 95\% containment band. Therefore, the inclusion of SNRs in the signal template can still result in relatively robust predictions for the sensitivity to axions. 

Given our axion decay emission templates, we have performed sensitivity estimates for existing or near-future radio telescopes including CHIME, HERA, CHORD, HIRAX, and BURSTT. To facilitate estimates for other compact survey interferometers, we have constructed a figure of merit that explains much of the variation in predicted sensitivity (not accounting for telescope position, which matters because much of our signal is concentrated on the Galactic plane). We find that arrays with a large \'etendue are best suited for axion DM searches of the kind considered in this work. Therefore, the goal of axion detection is well-aligned with other scientific objectives that benefit from a large \'etendue, for instance discovering FRBs or measuring the 21~cm power spectrum from the epoch of reionization and cosmic dawn. Thus, axion detection should be considered a collateral scientific objective of more futuristic arrays, for instance the proposed (full) Packed Ultra-wideband Mapping Array (PUMA)~\cite{CosmicVisions21cm:2018rfq}, which has a figure of merit that is a little more than five times larger than CHIME at 400 MHz. The ambitious proposals to measure the power spectrum from the cosmic dark ages on the far side of the moon~\cite{Burns:2021pkx,2022AAS...24020610P} would have a figure of merit that is approximately 20 times larger than HERA at 50~MHz; given the relevant frequency scalings, this could also potentially imply very strong sensitivity down to 100~kHz ($m_a\sim 10^{-9}$~eV), corresponding to the proposed frequency coverage.

We find that existing radio telescopes CHIME and HERA should already have impressive sensitivity to stimulated axion decay using archival data, potentially paving the way to a world-leading limit on axion DM. CHIME especially benefits from its large area and FOV. Our forecasted sensitivity lies in a region of axion parameter space that is highly complementary to other astrophysical probes, as well as terrestrial experiments. The radio searches for stimulated decay discussed in this work involve very different assumptions, parametric scalings, and systematic uncertainties compared to other axion searches, which will bolster any exclusions or claimed detections of axions. Additionally, the search strategy presented here has several advantages over other types of searches.

Terrestrial axion DM searches rely on the assumption that the DM is smoothly distributed with a local terrestrial density that is similar to the mean density of the Galactic neighborhood. This assumption may be substantially violated in axion cosmologies involving the early formation of axion mini-haloes~\cite{Buschmann:2019icd}, which would surive to the present day~\cite{Xiao:2021nkb, Shen:2022ltx} and impact the sensitivity of haloscope searches. In contrast, searches for axion decay stimulated by astrophysical sources are not dependent on the local terrestrial DM density. The aggregate all-sky signal strength comes from integrating the axion DM density over a deep column for many different lines of sight, and is therefore not affected by the presence of axion mini-haloes. 

Similar assumptions about the local DM distribution must be made for proposed setups involving stimulated decay induced by high-power terrestrial emitters~\cite{arza2019production,Arza:2021zqc,Arza:2022dng,Arza:2023rcs}. In these proposals, a megawatt-scale emitter produces a series of ``pulses'' lasting for a few hours, with each pulse at a different radio frequency. The search for the echo from stimulated decay can then constrain axion DM in the solar neighborhood. However, given the geometry depicted in Fig.~\ref{fig:gegenschein}, it is clear that the resulting signal is very sensitive to the local velocity distribution, particularly since the terrestrial emitter setup is far from the focused limit. The configuration of astrophysical sources, on the other hand, generally generates stimulated decay emission in the focused limit where the overall signal power is not affected by the DM velocity dispersion. 

Finally, other astrophysical constraints on axions involving compact objects (e.g. magnetic white dwarfs or pulsar polar caps) are strong and do not require an assumption that axions are the DM. However, the relevant signals for these searches typically scale as $\gagg^4$, while axion decay scales like $\gagg^2$. These scalings indicate that the path towards accessing lower values of $\gagg$ will be relatively less impeded for the stimulated decay signal. Moreover, many searches involving compact objects are limited by astrophysical uncertainties rather than instrumental sensitivity, whereas the search proposed here will significantly benefit from larger radio telescopes constructed in the future. 

The code used to produce the results in this paper is available \href{https://github.com/yitiansun/axion-mirror}{here}. A detailed analysis pipeline for CHIME will be presented in future work, which will pave the way towards the first use of this instrument to set limits on axion DM. We will additionally release an ensemble of signal templates that can be used by the radio-astronomy community to perform axion searches using other telescopes. 

\section*{Acknowledgements}
It is a pleasure to thank Manuel Buen-Abad, Cynthia Chiang, Matt Dobbs, JiJi Fan, Joshua Foster, Adrian Liu, Samar Safi-Harb, Tracy Slatyer, Chen Sun, Dallas Wulf, and Huangyu Xiao for useful conversations pertaining to this work.
YS was supported by the U.S. Department of Energy, Office of Science, Office of High Energy Physics of U.S. Department of Energy under grant Contract Number DE-SC0012567 through the Center for Theoretical Physics at MIT, and the National Science Foundation under Cooperative Agreement PHY-2019786 (The NSF AI Institute for Artificial Intelligence and Fundamental Interactions, \url{http://iaifi.org/}).
KS and HS acknowledge support from a Natural Sciences and Engineering Research Council of Canada (NSERC) Subatomic Physics Discovery Grant. KS additionally acknowledges support from the Canada Research Chairs program. HS was partially supported by a McGill Science Undergraduate Research Award. CL is supported by NASA through the NASA Hubble Fellowship grant HST-HF2-51536.001-A awarded by the Space Telescope Science Institute, which is operated by the Association of Universities for Research in Astronomy, Inc., under NASA contract NAS5-26555. KWM was supported by NSF grants (2008031, 2018490) and holds the Adam J. Burgasser Chair in Astrophysics. This analysis made use of \texttt{Numpy} \cite{harris2020array}, \texttt{Scipy} \cite{virtanen2020scipy}, \texttt{Astropy} \cite{robitaille2013astropy}, \texttt{Healpy} \cite{zonca2019healpy}, \texttt{Jax} \cite{jax2018github}, and \texttt{Matplotlib} \cite{hunter2007matplotlib}. 
\bibliographystyle{unsrt}
\bibliography{bib}
\appendix
\section{Detailed derivation of the stimulated decay intensity} \label{appd:Ist}
In this Appendix we present a more detailed derivation of the stimulated photon intensity in \eqref{eq:Ist}. To simplify the discussion, we ignore the time dependence of the source. We first start an expression for the stimulated decay flux of a infinitely far away source,
\begin{equation}
    S_\text{st}=\frac{\gagg^2}{16}S_{\nu,0}\int\rho(x_d)\dd x_d,
\end{equation}
where the integral occurs over the antipodal DM column \cite{ghosh2020axion}. We convert it to an expression of total stimulated emission power for DM occupying volume element,
\begin{equation}
    \begin{aligned}
        P_\text{st}(\hat i)&=\frac{\gagg^2}{16}P_\nu(\hat i) \rho(\vx_d) \dd x_{\hat i} \\
        &=\frac{\gagg^2}{16}S_\nu(\hat i) \dd A_{\perp\hat i} \rho(\vx_d) \dd x_{\hat i},
    \end{aligned}
\end{equation}
where $\dd x_{\hat i}$ corresponds to the {depth} of the volume element along the incoming photon direction and $\dd A_{\perp\hat i}$ is the perpendicular area element. Dividing out this volume element and writing the source flux in terms of source intensity and a small incoming photon solid angle $S_\nu=I_\nu\dd\Omega_i$, we obtain the DM emissivity
\begin{equation}
    j_\text{st}(\hat i)=\frac{\gagg^2}{16}\rho(\vx_d)I_\nu(\hat i)\dd\Omega_i.
\end{equation}
Applying the emitted photon angular distribution, we obtain the emissive intensity (emissivity per solid angle $\dd\Omega_e$) of the DM
\begin{equation}
    \frac{\dd j_\text{st}}{\dd\Omega_e}(\hat i, \hat e)=j_\text{st}(\hat i)\mathcal{F}(\hat i, \hat e).
\end{equation}
We consider the emission in a fixed direction $\hat e$, and we integrate over the incoming photon direction. Identifying $\hat i=\hx_{ds}$ in our setup, we have
\begin{equation}
    \begin{aligned}
        \frac{\dd j_\text{st}}{\dd\Omega_e}(\hat e)&=\frac{\gagg^2}{16}\int\dd\Omega_i\rho(\vx_d)I_\nu(\hat i)\mathcal{F}(\hat i, \hat e) \\
        &=\frac{\gagg^2}{16}\int\dd\Omega_{x_{ds}}\int\dd x_{ds}\rho(\vx_d)\frac{j_\nu(\vx_s)}{4\pi}\mathcal{F}(\hx_{ds}, \hat e) \\
        &=\frac{\gagg^2}{16}\int\dd^3\vx_s\rho(\vx_d)\frac{j_\nu(\vx_s)}{4\pi x_{ds}^2}\mathcal{F}(\hx_{ds}, \hat e).
    \end{aligned}
\end{equation}
In the final line, we have substituted $\dd^3\vx_{ds}=\dd^3\vx_s$ since we are keeping $\vx_d$ fixed. From the observer perspective, we idenity the emitted photon direction as $\hat e=-\hx_d$ so the stimulated decay intensity is
\begin{equation}
    \begin{aligned}
        I_\text{st}(-\hx_d)&=\int\dd x_d \frac{\dd j_g}{\dd\Omega_e}(-\hx_d) \\
        &=\frac{\gagg^2}{16}\int\dd x_d\dd^3\vx_s\rho(\vx_d)\frac{j_\nu(\vx_s)}{4\pi x_{ds}^2}\mathcal{F}(\hx_{ds}, -\hx_d).
    \end{aligned}
\end{equation}
Restoring the time dependence of the observed intensity due to the time dependence of the source emissivity and the time delay of the photons, we have
\begin{equation}
    \begin{aligned}
        I_\text{st}(-\hx_d,t)=&\frac{\gagg^2}{16}\iint\frac{\dd x_d\dd^3\vx_s}{4\pi x_{ds}^2}\rho(\vx_d) \\
        &j_\nu(\vx_s,t-x_d-x_{ds})\mathcal{F}(\hx_{ds}, -\hx_d).
    \end{aligned}
\end{equation}
\section{Dependence on modeling choices} \label{appd:model}
\subsection{Supernova Remnants} \label{appd:model-snr}
\begin{figure}[t]
    \centering
    \includegraphics[width=0.48\textwidth]{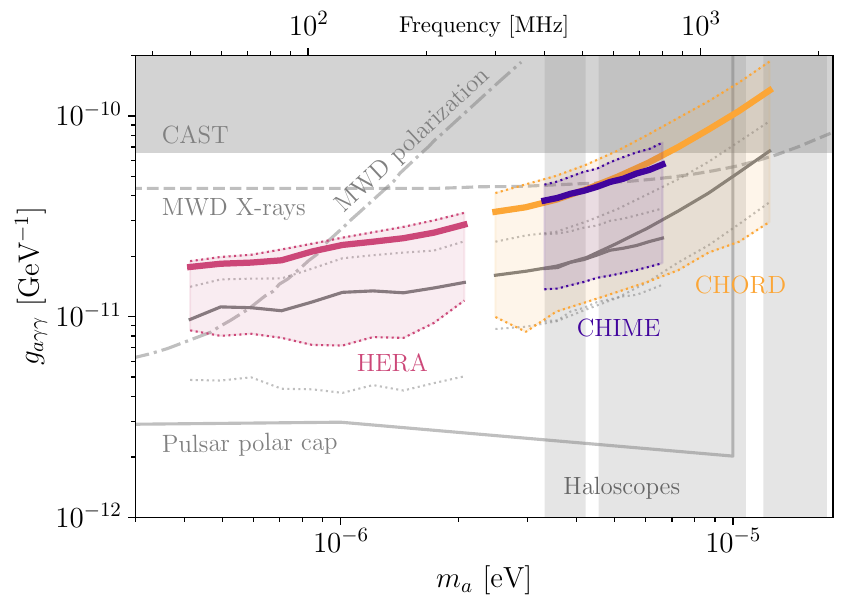}\vspace{-0.4cm}
    \caption{\textbf{Alternative electron model.} Projected reach of HERA, CHIME, and CHORD assuming an alternative electron model in the SNR adiabatic phase. This modeling choice impacts the luminosity histories of both the observed SNRs and the graveyard SNRs. For comparison, the reach and containment under the fiducial electron model are shown in grey. }
    \vspace{-0.3cm}
    \label{fig:reach_alte}
\end{figure}
In this Subsection, we consider the systematic uncertainty on the axion sensitivity due to SNR modeling choices. As previously discussed in Ref.~\cite{sun2022axion}, the modeling of the electron evolution during the Sedov-Taylor phase can affect the gegenschein signal strength. In addition to the classical model, we consider an adiabatic model where the product of the SNR volume $V$ and electron spectral normalization $K_e$ remain constant, $V\,K_e\sim\text{constant}$. \figref{fig:reach_alte} shows the 95\% containment of realizations of the total signal, assuming both the observed SNRs and graveyard SNRs follow the alternate electron model during the Sedov-Taylor phase. Note that we still sample the peak luminosity $L_\text{pk}$ and time $t_\text{pk}$ in the free expansion phase from the same distribution taken from \cite{bietenholz2021radio}. However, the conditions that $L_\text{pk}$ and $t_\text{pk}$ must satisfy for SNRs with observed fluxes change with the electron modeling choice (in order to reproduce the observed flux at the present day).

\begin{figure}[t!]
    \centering
\includegraphics[width=0.48\textwidth]{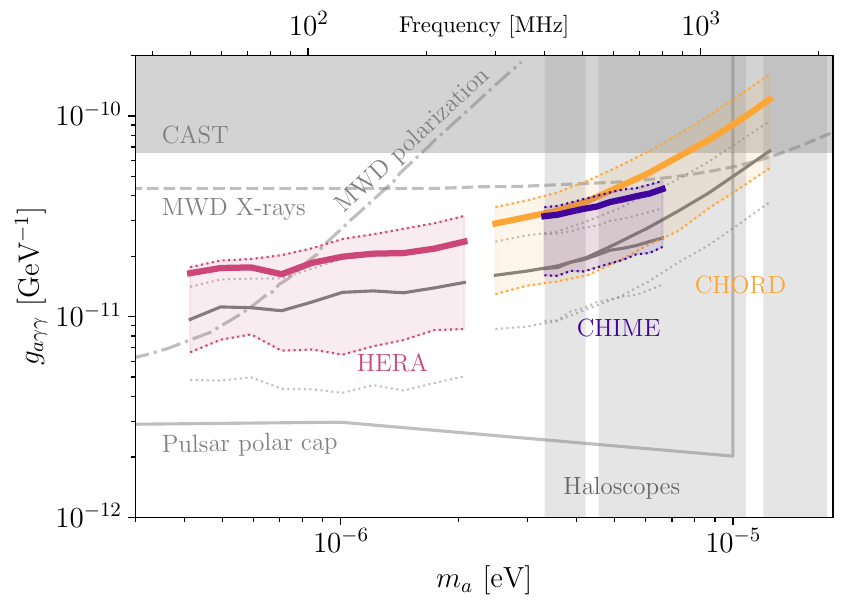}\vspace{-0.4cm}
    \caption{\textbf{Alternative free-expansion: constant luminosity.} Projected reach of HERA, CHIME, and CHORD assuming a constant radio luminosity in the free expansion phase ($<100$~years) for observed SNRs, and fiducial assumptions for graveyard SNRs. Given this change, the contributions from observed SNRs and graveyard SNRs become more similar, and both are contributing to the statistical spread. For comparison, the reach and containment under the fiducial free-expansion model are shown in grey. }
    \label{fig:reach_nofree}
    \vspace{-0.4cm}
\end{figure}

To further compare with Ref.~\cite{sun2022axion} and to highlight the importance of including the empirically motivated lightcurve in the free-expansion phase, in Fig.~\ref{fig:reach_nofree} we show the total sensitivity under the excessively conservative assumption that the radio luminosity stays constant in free expansion ($t<100$~years). In other words, when scaling back the lightcurve for observed SNRs from the present day, we assume that the luminosity does not rise past $t<100$~years.

\begin{figure}[t]
    \centering
    \includegraphics[width=0.48\textwidth]{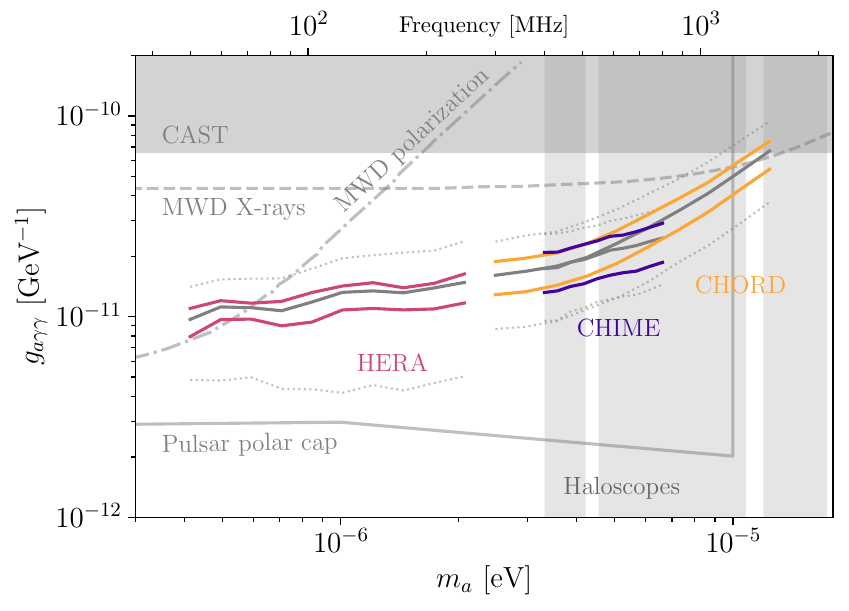}\vspace{-0.4cm}
    \caption{\textbf{Alternative free expansion-adiabatic transition time.} We show alternative transition times between phases of SNR evolution (for all SNRs) of 30~years and 300~years in comparison with the fiducial assumption of 100~years (grey). The assumption that the transition occurs 30 years after the SN produces a stronger constraint because the observed SNRs would have higher peak luminosities to reproduce the observed flux today. Therefore, under this assumption the observed SNRs would produce a stronger stimulating source for axion decay.}
    \label{fig:reach_tf}
    \vspace{-0.4cm}
\end{figure}
\begin{figure}[t]
    \centering
    \includegraphics[width=0.48\textwidth]{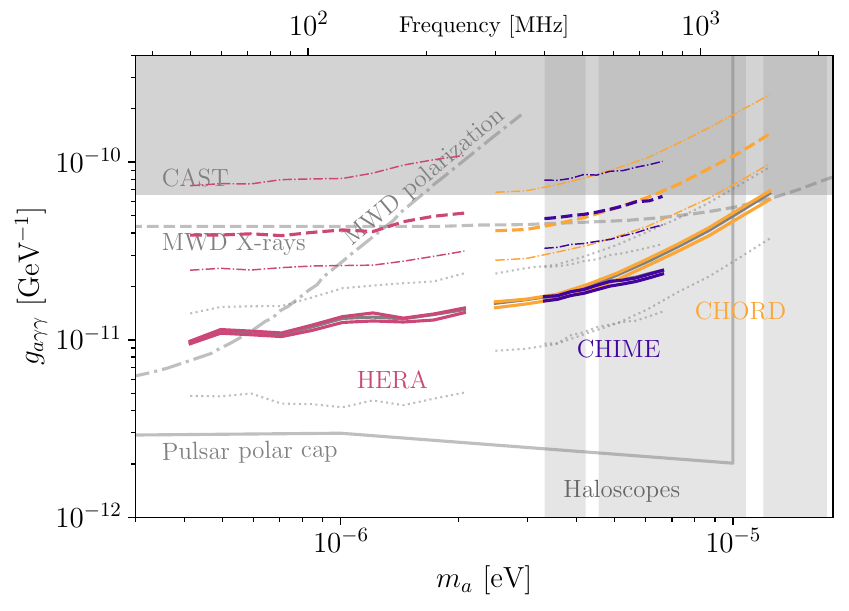}\vspace{-0.4cm}
    \caption{\textbf{Alternative galactic SNR rate.} We show the impact of alternative galactic SNR rate of 1.9/century and 12.0/century in comparison with the fiducial value of 4.6/century. The solid colored lines show the variation of the total sensitivity (higher rates yield better sensitivity). The dashed and dot-dashed lines show the graveyard-SNR-only contribution of the fiducial and alternative SNR rates, respectively.}
    \label{fig:reach_sr}
    \vspace{-0.4cm}
\end{figure}

In \figref{fig:reach_tf} and \figref{fig:reach_sr}, we show the effects of modeling choices of the transition time from free expansion phase to adiabatic phase, and the SNe rate in our Galaxy, respectively. For clarity, we only show the median reach for the alternative modeling choices, but they each still feature a statistical uncertainty band similar to the ones shown in \figref{fig:reach}. The difference in axion sensitivity caused by these modeling choices is small compared to the spread coming from different realizations of other more relevant SNR properties. Note that the SNR rate, while having a large effect on the SNR graveyard population, has only a small effect on the total sensitivity since the graveyard SNRs contribute less to the overall gegenschein signal.

\begin{figure}[t]
    \centering
    \includegraphics[width=0.48\textwidth]{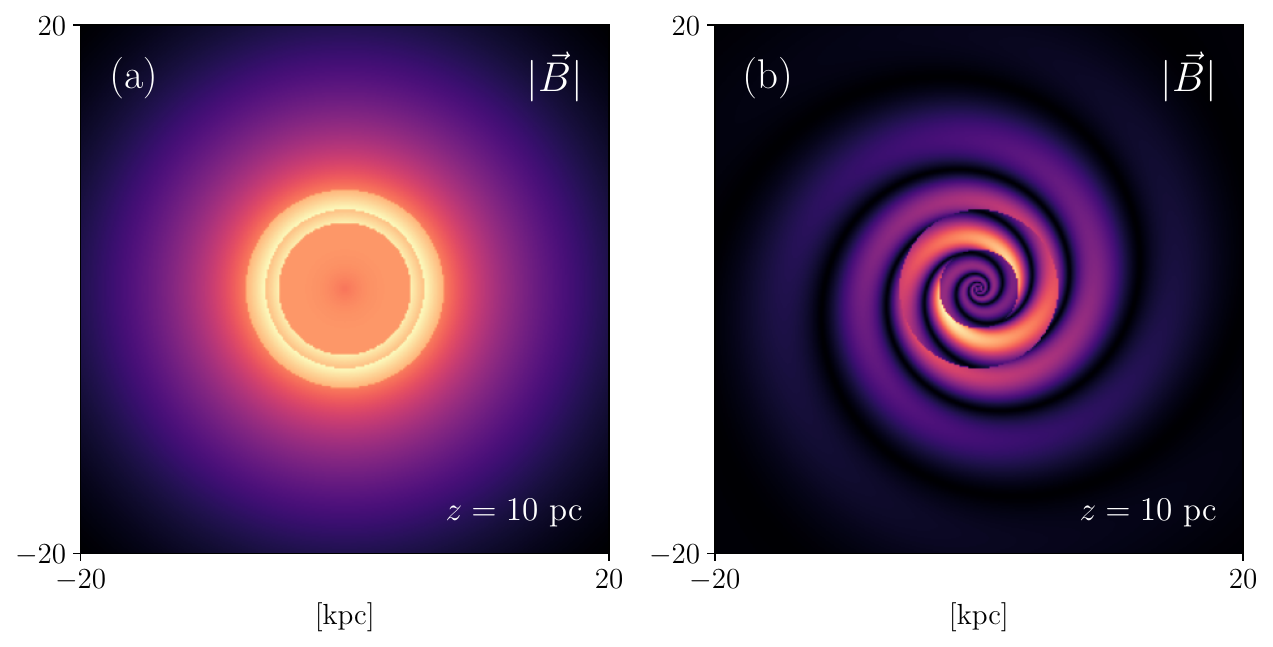}\vspace{-0.4cm}
    \caption{\textbf{Alternative galactic magnetic field models.} Here we show the cross sections of two alternative galactic magnetic field models presented in Ref. \cite{sun2008radio} for modeling the 3D distribution of the stimulating GSR source: (a) the Axisymmetric Spiral + Halo model and (b) Bisymmetric Spiral + Halo model.}
    \label{fig:gsr-B-models}
    \vspace{-0.4cm}
\end{figure}
\begin{figure}[t]
    \centering
    \includegraphics[width=0.48\textwidth]{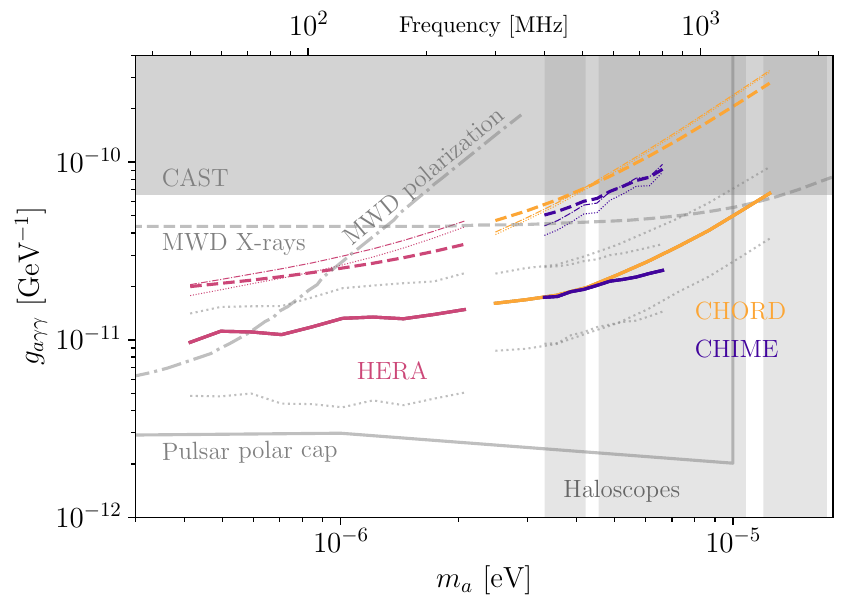}\vspace{-0.4cm}
    \caption{\textbf{Projected sensitivity for alternative galactic magnetic field models.} The colored thin dot-dashed and dotted lines represent the GSR-only contribution to the axion sensitivity from the Axisymmetric Spiral + Halo model and the Bisymmetric Spiral + Halo, respectively. The colored dashed lines represent the GSR-only sensitivity of the fiducial model of Ref.~\cite{jansson2012new}. The percent-level differences between the two alternative models and the fiducial model have been artificially inflated by a factor of 30 for visualization purposes. The colored solid lines represent the total sensitivity assuming the fiducial and alternate $B$-field models. The differences are not visible on this plot.}\vspace{-0.4cm}
    \label{fig:gsr-B-models-sens}
\end{figure}

\subsection{Galactic Synchrotron Radiation} \label{appd:model-gsr}
In \figref{fig:gsr-B-models}, we show the top-down view of two alternative galactic $B$-field models we use to construct a 3D GSR emissivity distribution. These models assume the same relativistic electron distribution NE2001 \cite{cordes2002ne2001} as our fiducial GSR model. As discussed in Section~\ref{sec:gsr}, we use 3D $B$-field and relativistic electron models to inform the LOS distribution of the GSR emissivity, and use the Haslam 408~MHz map to set the normalization of the integrated emissivity independently for each LOS. Therefore we expect the resulting axion stimulated decay signal to only depend weakly on the spatial modeling choice. In \figref{fig:gsr-B-models-sens}, we can see that the the axion reach is insensitive to the GSR model, even when only considering the subdominant GSR contribution to the overall signal. The predicted signal from the GSR therefore serves as a robust, model-independent lower bound on the total all-sky axion decay signal.

\section{Extended Results}
\begin{figure}[t]
    \centering
\includegraphics[width=0.48\textwidth]{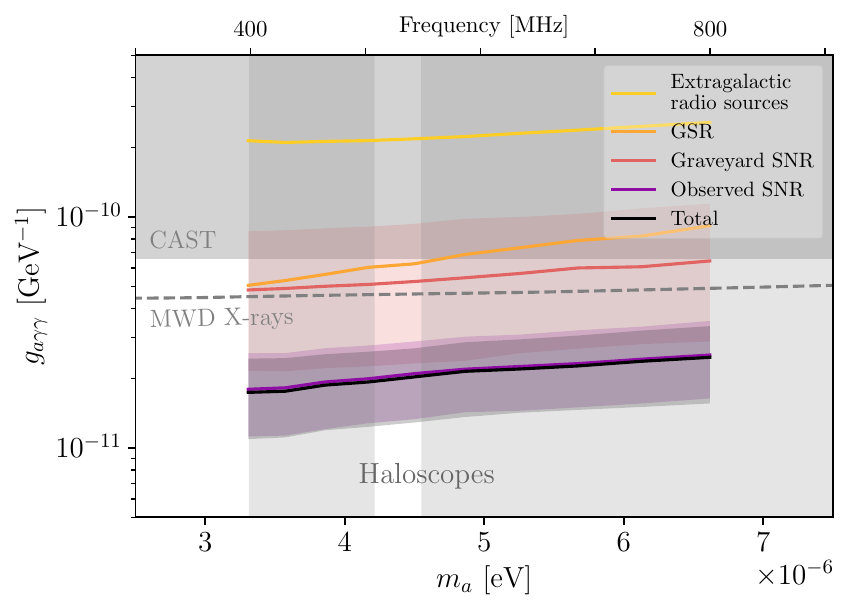}
\vspace{-0.8cm}
\caption{\textbf{Sensitivity contribution of various stimulating sources for CHIME.} The SNR sources have variable reach in different realizations of their properties; the median sensitivity is shown as a solid line with corresponding 95\% containment bands. The contributions from GSR and extragalactic sources are subject to much smaller systematic uncertainties, and thus they provide a robust upper (lower) limit to the overall reach in $\gagg$ (sensitivity).}
\vspace{-0.4cm}\label{fig:reach_CHIME_sources}
\end{figure}

\subsection{Comparison of various stimulating sources}
In \figref{fig:reach_CHIME_sources}, we compare the sensitivity achievable by CHIME if only a single type of stimulating source for axion decay is considered. The dominant contribution comes from the observed SNRs, followed by the graveyard SNRs, which have smaller fluxes since they tend to be farther away, which diminishes their ability to stimulate axion decay. Both populations of SNRs are characterized by systematic uncertainties due to incomplete information about the lightcurves where we can sample from empirically determined continuum distributions, as well as systematic uncertainties from binary modeling choices, as detailed in Appendix \ref{appd:model-snr}. The next-largest contributions to the stimulated decay signal come from GSR and extragalactic radio sources, which are subject to much smaller systematic uncertainties, as shown in Appendix \ref{appd:model-gsr} for the GSR component. These sources provide a robust minimum sensitivity to axion decay, making their inclusion important for the overall projection.

\subsection{Comparison of different instrumental configurations}
\begin{figure}[b!]
    \centering
\includegraphics[width=0.48\textwidth]{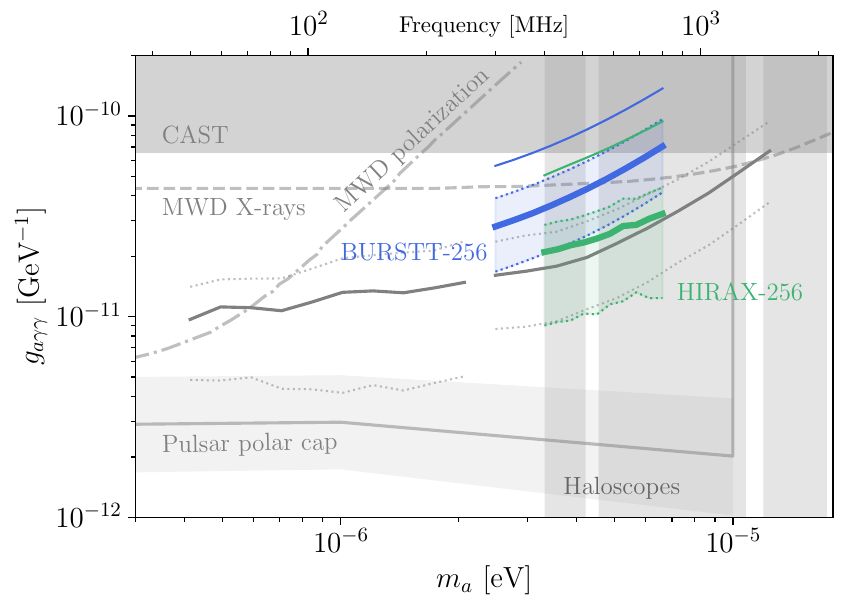}\vspace{-0.4cm}
    \caption{\textbf{Projected reach of HIRAX-256 and BURSTT-256.} Similar to \figref{fig:reach}, we show the projected sensitivity to the axion photon coupling $\gagg$ for HIRAX-256 and BURSTT-256, with HERA and CHORD shown for reference in grey. The bands represent 95\% containment of the statistical variation of 300 realizations of all stimulating sources, while the thin lines include only GSR and extragalactic radio sources.}
    \label{fig:reach_hb}
\end{figure}

In \figref{fig:reach}, we show the sensitivity of CHIME, CHORD, HERA, HIRAX-1024, and BURSTT-2048. In \figref{fig:reach_hb}, we show the same for HIRAX-256 and BURSTT-256, with the median sensitivity and 95\% coverage band of HERA and CHORD shown for reference. Of all the currently existing arrays, CHIME is one of the best in its frequency range thanks to its large collecting area and FOV. With $\sim$5 years of data already taken, CHIME is the ideal telescope to carry out the search for axion decay stimulated by astrophysical sources.

\end{document}